\documentclass[pre,amsmath,amssymb,twocolumn,superscriptaddress,
preprintnumbers,showpacs]{revtex4}

\allowdisplaybreaks

\usepackage{bm}
\usepackage{graphicx}
\usepackage{color}
\usepackage{enumerate}
\usepackage{tikz}
\usetikzlibrary{intersections, calc, through}
\def \mate<#1|#2|#3>{\mbox{$\langle {#1}|\,{#2}\,|{#3}\rangle$}}
\newcommand{\hsymb}[1]{\mbox{\strut\rlap{\smash{\Huge$#1$}}\quad}}

\newcommand\SC[4][\cellsize]{
\fill [#4, inner sep=0, outer sep=0] ($(2*#2*#1,-2*#3*#1)+(#1,#1)$) circle (#1);
\draw [inner sep=0, outer sep=0] ($(2*#2*#1,-2*#3*#1)+(#1,#1)$) circle (#1)
}

\newcommand\horDC[5][\cellsize]{
\fill [#4, inner sep=0, outer sep=0] ($(2*#2*#1,-2*#3*#1)+(#1,2*#1)$) arc [start angle=90, end angle=270, radius=#1];
\fill [#5, inner sep=0, outer sep=0] ($(2*#2*#1,-2*#3*#1)+(#1,0)$) arc [start angle=-90, end angle=90, radius=#1];
\draw [inner sep=0, outer sep=0] ($(2*#2*#1,-2*#3*#1)+(#1,#1)$) circle (#1)
}
\newcommand\verDC[5][\cellsize]{
\fill [#4, inner sep=0, outer sep=0] ($(2*#2*#1,-2*#3*#1)+(2*#1,#1)$) arc [start angle=0, end angle=180, radius=#1];
\fill [#5, inner sep=0, outer sep=0] ($(2*#2*#1,-2*#3*#1)+(0,#1)$) arc [start angle=180, end angle=360, radius=#1];
\draw [inner sep=0, outer sep=0] ($(2*#2*#1,-2*#3*#1)+(#1,#1)$) circle (#1)
}
\newcommand{\bSC}[1][\cellsize]{\tikz[baseline=0.5pt]\SC[#1]{0}{0}{blue!80!cyan};}
\newcommand{\uSC}[1][\cellsize]{\tikz[baseline=0.5pt]\SC[#1]{0}{0}{black};}
\newcommand{\gSC}[1][\cellsize]{\tikz[baseline=0.5pt]\SC[#1]{0}{0}{green};}
\newcommand{\rSC}[1][\cellsize]{\tikz[baseline=0.5pt]\SC[#1]{0}{0}{red};}
\newcommand{\hrgDC}[1][\cellsize]{\tikz[baseline=0.5pt]\horDC[#1]{0}{0}{red}{green};}
\newcommand{\hgrDC}[1][\cellsize]{\tikz[baseline=0.5pt]\horDC[#1]{0}{0}{green}{red};}
\newcommand{\vrgDC}[1][\cellsize]{\tikz[baseline=0.5pt]\verDC[#1]{0}{0}{red}{green};}
\newcommand{\vgrDC}[1][\cellsize]{\tikz[baseline=0.5pt]\verDC[#1]{0}{0}{green}{red};}

\newcommand{\ZlineA}{\uSC\hgrDC\bSC\hrgDC}
\newcommand{\ZlineB}{\bSC\hrgDC\uSC\hgrDC}
\newcommand{\ZlineC}{\hgrDC\bSC\hrgDC\uSC}
\newcommand{\ZlineD}{\hrgDC\uSC\hgrDC\bSC}
\newcommand{\ZlineE}{\bSC\uSC\bSC\uSC}
\newcommand{\ZlineF}{\vrgDC\vgrDC\vrgDC\vgrDC}
\newcommand{\ZlineG}{\uSC\bSC\uSC\bSC}
\newcommand{\ZlineH}{\vgrDC\vrgDC\vgrDC\vrgDC}

\newcommand{\ZlineAhalf}{\uSC\hgrDC}
\newcommand{\ZlineBhalf}{\bSC\hrgDC}

\newcommand{\ZlineEhalf}{\bSC\uSC}
\newcommand{\ZlineFhalf}{\vrgDC\vgrDC}
\newcommand{\ZlineGhalf}{\uSC\bSC}
\newcommand{\ZlineHhalf}{\vgrDC\vrgDC}
\newcommand{\oo}{\mathcal{O}} % order
\newcommand{\f}[2]{{\frac{#1}{#2}}} % shortened \frac
\newcommand{\s}[1]{\sqrt{#1}}       % shortened \sqrt
\def\l{\left}  % shortened \left
\def\r{\right} % shortened \right
\newcommand{\fgref}[1]{{\figurename\;\ref{#1}}} % ref of figure
\newcommand{\supsecref}[1]{{\appendixname\;\ref{#1}}}
\newcommand{\ket}[1]{{|{#1}\rangle}}
\newcommand{\bra}[1]{{\langle{#1}|}}
\newcommand{\braket}[2]{{\langle{#1}|{#2}\rangle}}
\newcommand{\abs}[1]{{\left|#1\right|}}  % absolute value
\newcommand{\cE}{\mathcal{E}} % energy
\newcommand{\cM}{\mathcal{M}} % manifold
\newcommand{\rad}{*}
\newcommand{\cir}{\odot}

\makeatletter
\newcommand{\customlabel}[2]{%
\protected@write \@auxout {}{\string \newlabel {#1}{{#2}{}}}}
\makeatother

\newcommand{\papertitle}{{Dynamical Pattern Selection of Growing Cellular Mosaic in Fish Retina}}

\newcommand{\QHP}{{Theoretical Research Division, RIKEN Nishina Center, Saitama 351-0198, Japan}}
\newcommand{\TB}{{Theoretical Biology Laboratory, RIKEN, Saitama 351-0198, Japan}}
\newcommand{\iTHEMS}{Interdisciplinary Theoretical and Mathematical Sciences (iTHEMS), RIKEN, Saitama 351-0198, Japan}
\newcommand{\iTHES}{{Interdisciplinary Theoretical Science Research Group (iTHES), RIKEN, Saitama 351-0198, Japan}}
\newcommand{\CREST}{{CREST, JST, Kawaguchi 332-0012, Japan}}

\begin{document}

\title{\papertitle}

\affiliation{\QHP}
\affiliation{\TB}
\affiliation{\iTHEMS}
\affiliation{\iTHES}
\affiliation{\CREST}
\author{Noriaki \surname{Ogawa}}
\email{noriaki@riken.jp}
\affiliation{\QHP}
\affiliation{\iTHES}
\author{Tetsuo \surname{Hatsuda}}
\email{thatsuda@riken.jp}
\affiliation{\QHP}
\affiliation{\iTHEMS}
\affiliation{\iTHES}
\author{Atsushi \surname{Mochizuki}}
\email{mochi@riken.jp}
\affiliation{\TB}
\affiliation{\iTHEMS}
\affiliation{\iTHES}
\affiliation{\CREST}
\author{Masashi \surname{Tachikawa}}
\email{mtach@riken.jp}
\affiliation{\TB}
\affiliation{\iTHES}

\begin{abstract}

A Markovian lattice model for photoreceptor cells
is introduced to describe the growth of mosaic patterns on fish retina.
The radial stripe pattern observed in wild-type zebrafish
is shown to be selected naturally during the retina growth,
against the geometrically equivalent, circular stripe pattern.
The mechanism of such dynamical pattern selection 
is clarified on the basis of 
both numerical simulations and theoretical analyses,
which find that the successive emergence of local defects plays 
a critical role in the realization of the wild-type pattern.

\end{abstract}

\preprint{RIKEN-QHP-248, RIKEN-STAMP-29}

\maketitle

%------------------------------
\section{Introduction}
%------------------------------

Arrangement of the color detecting photoreceptors (cone cells)
on a retina is believed to be functionally important for animals to determine the spatial
resolution of the color imaging \cite{Cook-Eglen}.
In some teleost fishes, the cone cells sensitive to
green (G \gSC), red (R \rSC),  blue (B \bSC)  and ultra-violet (UV \uSC) lights
with G and R forming a pair called the ``double cones'' (GR \hgrDC),
form two-dimensional patterns:  They are called the ``cone mosaics'' and
have been known since 19th century (see e.g. \cite{Lyall:1957} and references therein).
However, the mechanism for the pattern formation is still unclear.

\begin{figure}
\vspace{-7pt}
\includegraphics[width=0.48\textwidth]{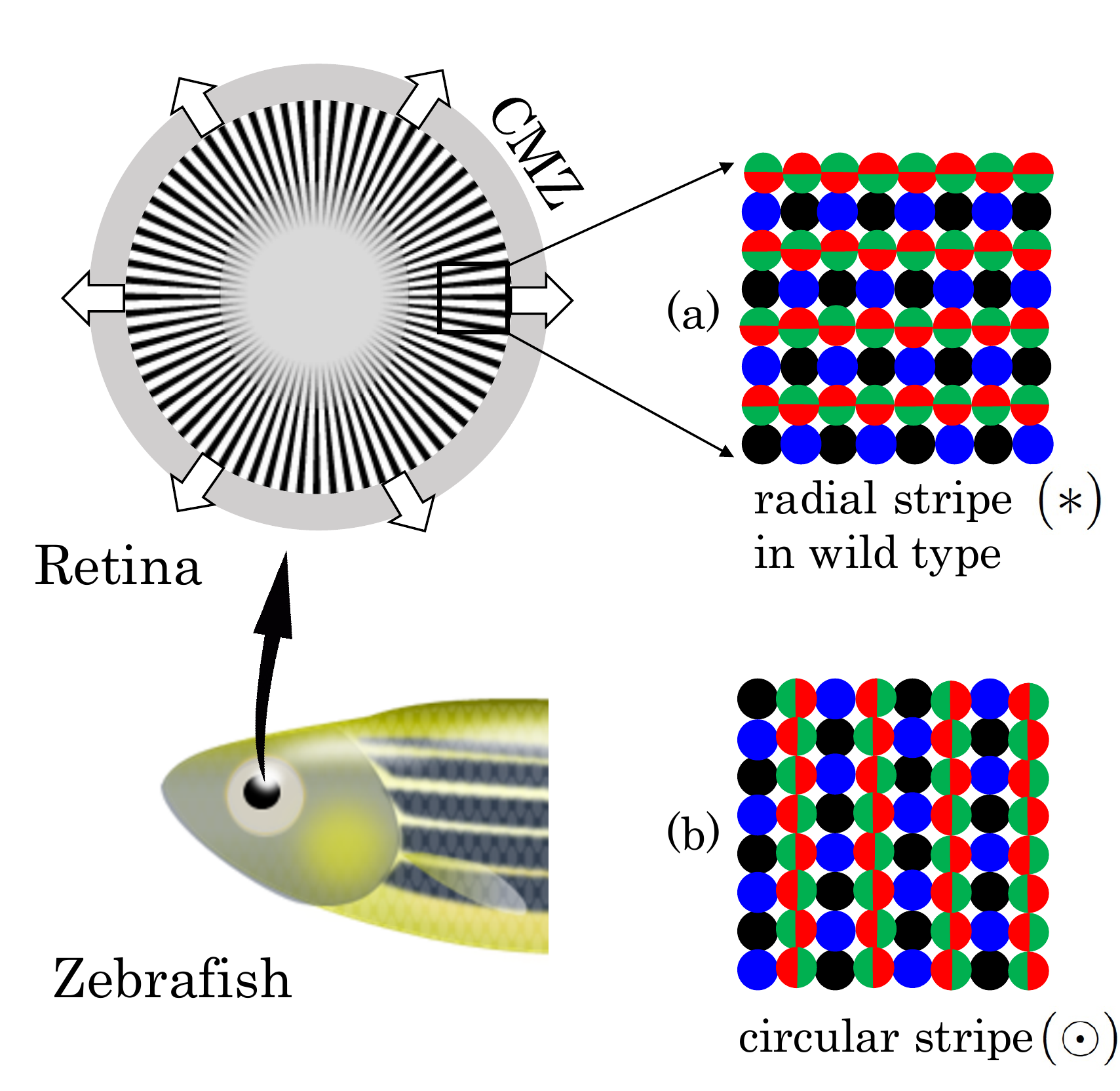}
\vspace{-15pt}
\caption{Schematic picture of the Zebrafish retina and possible cone mosaic patterns.
Typical size of the retina and the cone cell are
several hundred $\mu$m (3-4 days after fertilization)
and a few $\mu$m (along the retina surface), respectively.
The ciliary marginal zone (CMZ) where new cone cells are born
is denoted by gray area surrounding the retina.
}
\label{fig:Patterns}
\end{figure}

Among others, the cone mosaic of the zebrafish shown in \fgref{fig:Patterns}
has a prominent feature with two major characteristics;
directionality and long periodicity \cite{Fadool:2008}.
Stripes of cells run radially from the central to the marginal region of the retina, and the periodicity of the
pattern differs between radial and circular directions as shown in \fgref{fig:Patterns}(a).
Although the two patterns, (a) and its +90$^{\circ}$ rotation (b), have the same \textit{c2mm} symmetry of the wallpaper group
\cite{Doris:1978} and thus the same binding energy,
only pattern (a) is realized in nature.
In this paper, we refer to (a) as the ``radial'' ($\rad$) stripe pattern
since it is along the growing direction of the retina,
and (b) as the ``circular'' ($\cir$) one.

The retina of an adult fish grows continuously throughout its life by generating new photoreceptors
in the germinal zone of the retinal margin called the ciliary marginal zone (CMZ).
The rearrangement of those new cone cells
{\em after} cellular differentiation is
considered to play a crucial role in mosaic pattern formation.
This is supported by a cell-trace experiment
in which most of the cells are found not to change their subtypes
after the first determination \cite{Suzuki:2013}.

Theoretically, cell rearrangement models for cone mosaics have been formulated
on a two-dimensional square lattice:
It was found that
different patterns (such as the zebrafish pattern with \textit{c2mm} symmetry
and the medaka pattern with \textit{p4mg} symmetry) can be generated successfully
by stochastic cell movements biased by appropriate cell-cell adhesion force
\cite{Tohya:1999,Mochizuki:2002,Tohya:2003}.
However, such models have failed to distinguish the characteristics of directionality,
such as the distinction between (a) and (b) in \fgref{fig:Patterns}.

The main purpose of this paper is to propose a mechanism in which
the sequential accretion of new cone cells from CMZ triggers
dynamical pattern selection (DPS) among those with the same binding energy.
We demonstrate it explicitly by taking the zebrafish as a characteristic example.
Although we will focus on the cellular mosaic patterns in this paper,
they have suggestive similarities with other examples
in non-equilibrium physics
of pattern formations in growing systems,
%such as the crystal growth, patterns in granular materials, copolymerization etc,
where various spatial or temporal patterns emerge due to infinitesimally small perturbations \cite{GL1999}.

This paper is organized as follows.
In Sec.\ref{sec:model},
we introduce a Markovian  model  with a transition matrix $\hat{T}$ for the retinal growth.
In Sec.\ref{sec:t=0}, we classify
possible stationary patterns obtained from our model.
In Sec.\ref{sec:t>0},
we investigate the dynamical pattern formation by numerical simulations and by theoretical analyses
 on  $\hat{T}$, and we introduce the concept of dynamical pattern selection.
Sec.\ref{sec:discussion} is devoted to summary as well as discussions on future directions.
Details of the theoretical analyses on DPS in the full model and in a minimal toy-model are
 discussed in \supsecref{sup:perturbation}
and \supsecref{sup:numerical}, respectively.

%-----  Fig1 (fish retina) -------------
%----------------------------------------

%------------------------------
\section{Model for Pattern Formation}
\label{sec:model}
%-----------------------------

%-----  Fig2 (lattice model) -------------
\begin{figure}
  \centering
  \includegraphics[width=0.47\textwidth]{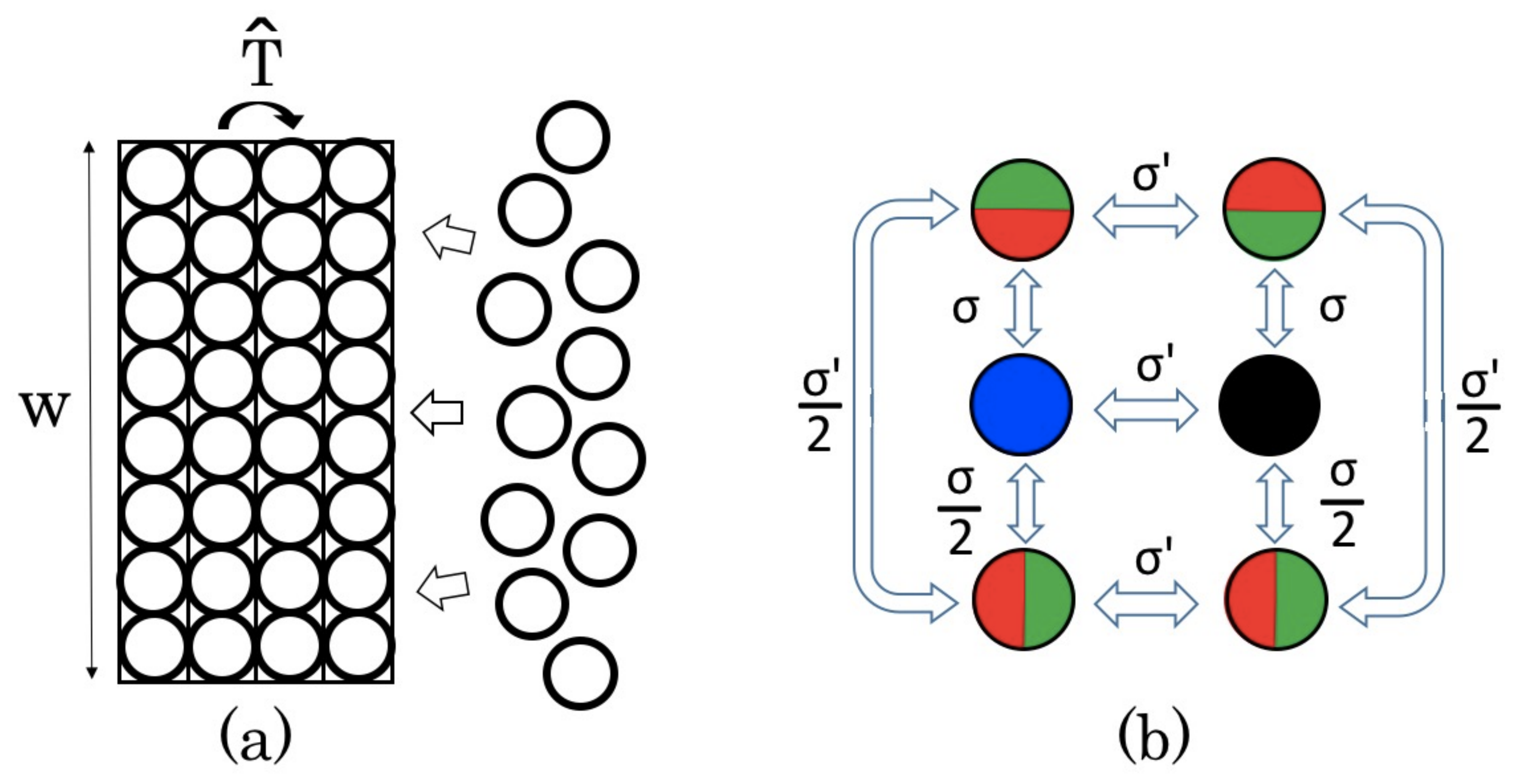}
\vspace{-5pt}
  \caption{(a)
  Radial growth of a part of the fish retina with a vertical size $w$
  due to the  accretion of new cone cells.
  The transition matrix from one column to the next is denoted by $\hat{T}$
  (see \eqref{eq:Markov}).
  (b) Binding energies between neighboring sites for Zebrafish patterns.}
  \label{fig:markov}
\end{figure}
%-----------------------------------------------

\subsection{Static Lattice Model}
Before introducing our dynamical model for retina growth,
we briefly review a static lattice model for mosaic patterns originally 
introduced by one of the present authors \cite{Tohya:1999}.
Let us consider a two-dimensional square lattice without growth
 and assume that each of the lattice sites is
occupied either by 
$\bSC$, $\uSC$, $\hgrDC$, $\vgrDC$, $\hrgDC$ or $\vrgDC$.
To quantify the effective ``\textit{binding energy}'' 
between neighboring cells, the parameters $\sigma_{pq}$
with $p$ and $q$ taking either B, R, G, or UV are introduced. 
The parameters $\sigma_{pq}$ could be related to
the intercellular adhesion
by the binding proteins on the cone cell membranes,
although specific proteins are not fully identified experimentally.
The binding energies to the double cones are assumed to be proportional to the contact area,
e.g. the side-by-side coupling
between $\bSC$ and $\vgrDC$ is chosen to be $(\sigma_\mathrm{_{BG}}+\sigma_\mathrm{_{BR}})/2$.
The total effective energy of each cell arrangement is defined 
as the sum of  binding energies of all the neighboring pairs: It
determines possible stable patterns  under the chosen parameter set.

Analyses of the ``phase diagram'' of cell patterns in the
multi-dimensional $\sigma_{pq}$-space
\cite{Tohya:1999,Mochizuki:2002},
showed that 
the patterns (a)(b) in \fgref{fig:Patterns}
are equally realized as stable states
if the three inequalities below are simultaneously satisfied
\footnote{
For other cases  such as the  Medaka pattern,
$\sigma_{pq}$ with $p,q$ taken not only for
neighboring cells but also for the diagonal cells is known to be necessary \cite{Tohya:2003}.}:
\begin{subequations}
\label{eq:TMIzebracond}
\begin{align}
3\sigma_\mathrm{_{RG}} - \sigma_\mathrm{_{BU}} &< \sigma_\mathrm{_{BR}} + \sigma_\mathrm{_{UG}},\\
 3\sigma_\mathrm{_{BU}} - \sigma_\mathrm{_{RG}} &< \sigma_\mathrm{_{BR}} + \sigma_\mathrm{_{UG}},\\
 \sigma_\mathrm{_{BR}}+\sigma_\mathrm{_{UG}}
 &< 2(\sigma_\mathrm{_{BU}}+\sigma_\mathrm{_{RG}}) - (\sigma_\mathrm{_{BG}}+\sigma_\mathrm{_{UR}}).
\end{align}
\end{subequations}

\subsection{Markovian Lattice Model}

To take into account the fact that a retina grows dynamically by receiving
new cone cells born in CMZ, 
we extract a rectangular region of the retina margin
as indicated in \fgref{fig:Patterns} and 
 introduce a dynamical  model
where cone cells are supplied stochastically 
from the ``CMZ pool'' to the two-dimensional rectangular  lattice  column-by-column
as shown in \fgref{fig:markov}(a).
The radial growth of the retina is represented by the horizontal growth toward the right direction,
while the vertical size of the lattice is fixed to be $w$
\footnote{%
In realistic radial growth, $w$ is not a constant and gradually increases.
It leads to dislocations of the cells in the mosaic and insertions of additional stripes.
We do not consider such effects in this paper.
}%
.

The growth is dictated by the action of a transition matrix $\hat{T}$ whose explicit form will be specified shortly.
In the CMZ pool, we consider B, UV and the double cone (RG) with the number ratio $1:1:2$,
anticipating the  possible stripe structures in \fgref{fig:Patterns}(a,b).
Then, we have six degrees of freedom on each lattice site with the  probability,
$
\bSC:\uSC:\hgrDC:\vgrDC:\hrgDC:\vrgDC
= \frac{1}{4} : \frac{1}{4} : \frac{1}{8} : \frac{1}{8} : \frac{1}{8} : \frac{1}{8}
$.

For the choice of binding parameters $\sigma_{pq}$,
we take the inequalities \eqref{eq:TMIzebracond} as guidelines:
our particular choice of $\sigma_{pq}$ is shown in \fgref{fig:markov}(b):
\begin{align}
\label{eq:paramrange}
&\sigma \equiv \sigma_\mathrm{_{BR}}=\sigma_\mathrm{_{UG}},
\quad
\sigma' \equiv \sigma_\mathrm{_{BU}}=\sigma_\mathrm{_{RG}}
\quad
(1 < \sigma/\sigma' < 2),
\nonumber\\
&\sigma_\mathrm{_{BG,UR,BB,RR,GG,UU}}=0.
\end{align}
Therefore, in \fgref{fig:markov}(b),
we have $\sigma'/2 < \sigma/2 < \sigma' < \sigma$.
In our analyses below, we take a specific value
\begin{align}
\sigma/\sigma'=3/2.
\end{align}
We have checked that the essential conclusions of our paper are unchanged as long as $1 < \sigma/\sigma' < 2$.

In the Markov chain shown in \fgref{fig:markov}(a),
the lattice sites are empty at the beginning and the
six types of cone cells in \fgref{fig:markov}(b) occupy the sites
column-by-column in a stochastic manner.
Each column with a periodic boundary condition in the vertical direction has
6$^w$ different configurations labeled by the index $j (=1, 2, \cdots , N=6^w)$.
The probability of having the configuration $j$ in the $n$-th column is defined as
$P_j^{(n)}$ so that the probability vector for the $n$-th column reads
$\ket{P^{(n)}}\equiv (P_1^{(n)}, P_2^{(n)}, \cdots , P_{N}^{(n)})^\top$ with
$\sum_j P_j^{(n)}=1$,
where $(\cdots)^\top$ denotes the transpose ($-90^{\circ}$ rotation).
The Markov chain is generated by the $N \times N$ transition matrix  $\hat{T}$ as
\begin{align}
\label{eq:Markov}
  \ket{P^{(n+1)}} = \hat{T}\ket{P^{(n)}},
\end{align}
where the matrix element $T_{ij}$ with  $\sum_{i=1}^{N} T_{ij}=1$ is
the probability of  having a configuration $i$ in the $(n+1)$-th column
after the configuration $j$ in the $n$-th column.

To model the retinal growth,
we consider $T_{ij}$  having a Boltzmann form
with an ``effective'' temperature $\beta^{-1}$:
\begin{align}
\label{eq:Pxy}
  {T}_{ij} = \f{p_i e^{-\beta \cE_{ij}}}{N_j},
\quad
  N_j = \sum_{i}{p_i e^{-\beta \cE_{ij}}},
\nonumber\\
  \cE_{ij}(\sigma,\sigma') = - B^\mathrm{(intra)}_{i} - B^\mathrm{(inter)}_{ij} \le 0.
\end{align}
Here $B^\mathrm{(intra)}_{i}$ is a sum of the intra binding energies in the configuration $i$,
while $B^\mathrm{(inter)}_{ij}$ is a sum of the inter binding energies between $i$ and $j$.
The prior probability factor $p_i$ is computed by taking into account the number ratio of
the cone cells in the CMZ pool:
For example, $p_i=\f{1}{4}\times\f{1}{8}\times\f{1}{4}\times\f{1}{8}$ for $i=(\ZlineA)^\top$.
Magnitude of the fluctuation among  cell patterns with different binding energies  is  dictated by $\beta^{-1}$.
Since $\cE_{ij}$ is a linear function of $\sigma$ and $\sigma'$,
$\beta \cE_{ij}$ and $T_{ij}$ can be written in terms of
the dimensionless variables,
$t \equiv {2}/{\beta \sigma'}$ (reduced temperature) and
$\sigma/\sigma'$ (coupling ratio).
By definition, the fluctuation becomes large as $t$ increases.

For later convenience, we introduce a bra-ket notation:
Each configuration $i$ is represented by a state vector $\ket{i}$,
so that $T_{ij} = \bra{i}\hat{T}\ket{j}$.
The eigenvalues of $\hat{T}$ and the corresponding left-right
eigenvectors are defined by
\begin{align}
\bra{\lambda_l}\hat{T} &=  \bra{\lambda_l}\lambda_l,
\quad
\hat{T}\ket{\lambda_l}  = \lambda_l\ket{\lambda_l},
\quad
(l=1, 2, \cdots, N)
\end{align}
with a relative normalization,
$\braket{\lambda_l}{\lambda_{l'}}  = \delta_{l l'}$.
Then, we have
\begin{align}
\label{eq:T-P}
\hat{T}^n= \sum_l \ket{\lambda_l} \lambda_l^n \bra{\lambda_l},
\quad
 \ket{P^{(n)}} = \sum_l c_l \lambda_l^n \ket{\lambda_l},
\end{align}
where  $c_l$'s are the expansion coefficients at $n=0$.
Note that $\lambda_l$'s are in general complex, since $\hat{T}$ is non-symmetric.
 
At $t=0$, $\hat{T}$ is reducible, i.e.,
there appear multiple closed subspaces and $\ket{P^{(\infty)}}$ is not unique.
Correspondingly, there arise multiple eigenvectors with the eigenvalue $1$.
For $t > 0$, all the components of $\hat{T}$ are positive,
so that the finite Markov chain becomes irreducible and aperiodic \cite{OH:2002}.
Then the Perron-Frobenius theorem for positive matrices leads to the facts that:
\begin{enumerate}[(i)]
\item
\label{item:lambda1}
$\hat{T}$ has one and only one unit eigenvalue $\lambda_{l=1}=1$
with the left eigenvector being $\bra{\lambda_1} =(1,1,1,\cdots)$.
\item
The absolute values of other eigenvalues obey $|\lambda_l| < 1$
with the sum of components for each right eigenvectors being $0$ due to $\braket{\lambda_1}{\lambda_{l'}} =0$ ($l\ne l'$).
\end{enumerate}
Consequently, there exists a unique stationary distribution vector $\ket{P^{(\infty)}}$ (which is nothing but $\ket{\lambda_1}$),
irrespective of the initial vector.
Furthermore the property (\ref{item:lambda1}) and 
$\sum_j P_j^{(n)}=1$ imply $\braket{\lambda_1}{P^{(n)}}=1$,
which in turn leads to $c_1=1$ by \eqref{eq:T-P}.
The essential question in DPS is to identify
a particular pattern which dominates
the state $\ket{\lambda_1}$ at $n\rightarrow \infty$.

%-----------------------------------------------
\section{Stationary patterns at $t=0$}
\label{sec:t=0}
%-----------------------------------------------

For zero effective temperature,
the following 8 columns play crucial roles in pattern formation:
(A) $(\ZlineA\cdots)^\top$, (B) $(\ZlineB\cdots)^\top$, (C) $(\ZlineC\cdots)^\top$,
(D) $(\ZlineD\cdots)^\top$, (E) $(\ZlineE\cdots)^\top$, (F) $(\ZlineF\cdots)^\top$,
(G) $(\ZlineG\cdots)^\top$, (H) $(\ZlineH\cdots)^\top$,
where $\cdots$ denotes the repetition.
In terms of the bra-ket notation, $\ket{A} = (1,0,0,0, \cdots)^\top$,
$\ket{B} = (0,1,0,0, \cdots)^\top$,  $\ket{C} = (0,0,1,0 \cdots)^\top$, etc.
Then, $\hat{T}_0$ (the transition matrix at $t=0$) in the state-space spanned by
$\cM\equiv\{A, B, C, D, E, F, G, H \}$
and the others are reducible and sparse:
\vspace{-10pt}
\begin{align}
\label{eq:T0}
  \hat{T}_0 =&
  \begin{array}{l}
\text{\tiny\mbox{$
\quad\quad
      A \;\; B \;\;\, C \;\; D \;\;\, E \;\;\, F \;\;\, G \;\; H \;\; \cdots
$}}
\vspace{4pt}
\\
\text{\footnotesize\mbox{$
    \left(
    \begin{array}{c|l}
    \begin{array}{cccc|cccc}
    0 & 1 & 0  & 0  &   &   &   &   \\
    1 & 0 & 0  & 0  &   &   &   &   \\
    0 & 0 & 0  & 1  &   & \multicolumn{2}{c}{\hsymb{0}}  &  \\
    0 & 0 & 1  & 0  &   &   &   &   \\
\hline
      &   &   &   & 0 & 0 & 0 & 1   \\
      &   &   &   & 1 & 0 & 0 & 0   \\
      & \multicolumn{2}{c}{\hsymb{0}}  &   & 0 & 1 & 0 & 0 \\
      &   &   &   & 0 & 0 & 1 & 0 \\
    \end{array}
      & \hsymb{\cdots}\\
      \hline
      & \hspace{25pt}\\
     \hsymb{0} 
      & \hsymb{\ddots}
\\
    \end{array}
    \right)
$}}
  \end{array}.
\end{align}

Since $(\hat{T}_0)_{i \not\in \cM,j\in\cM} =0$,
all the probability flows into the subspace $\cM$,
while it does not flow out from $\cM$.
Furthermore, the subspace $\cM$ is reducible to three closed sets, 
$\{ A,B \}$, $\{ C,D \}$, and  $\{ E,F,G,H\}$, so that they form 
the following stable patterns;
\begin{align}
\label{eq:w=4ZebraPatterns}
&\xrightarrow{\text{\; Growth \;}}
\;
\nonumber\\[-8pt]
&\begin{array}{ccccc}
  \begin{array}{l}
    {\scriptstyle A\,B\,A\,B\,A\,B\, \cdots}\\[-2pt]
    \ZlineG\ZlineGhalf \\[-6.5pt] 
    \ZlineH\ZlineHhalf \\[-6.5pt] 
    \ZlineE\ZlineEhalf \;\; \cdots \\[-6.5pt] 
    \ZlineF\ZlineFhalf \\[-6.5pt] 
    \ZlineG\ZlineGhalf \\[-6.5pt] 
    \ZlineH\ZlineHhalf \\[-6pt] 
    \;\vdots\;\;\vdots\;\,\vdots\;\;\vdots\;\,\vdots\;\,\vdots
  \end{array},
  &&
  \begin{array}{l}
    {\scriptstyle C\,D\,C\,D\,C\,D\, \cdots}\\[-2pt]
    \ZlineH\ZlineHhalf \\[-6.5pt] 
    \ZlineE\ZlineEhalf \\[-6.5pt] 
    \ZlineF\ZlineFhalf \;\; \cdots \\[-6.5pt] 
    \ZlineG\ZlineGhalf \\[-6.5pt] 
    \ZlineH\ZlineHhalf \\[-6.5pt] 
    \ZlineG\ZlineGhalf \\[-6pt] 
    \;\vdots\;\;\vdots\;\,\vdots\;\;\vdots\;\,\vdots\;\,\vdots
  \end{array},
  &&
  \begin{array}{l}
    {\scriptstyle E\,F\,G\,H\,E\,F\, \cdots}\\[-2pt]
    \ZlineB\ZlineBhalf \\[-6.5pt] 
    \ZlineA\ZlineAhalf \\[-6.5pt] 
    \ZlineB\ZlineBhalf \;\; \cdots \\[-6.5pt] 
    \ZlineA\ZlineAhalf \\[-6.5pt] 
    \ZlineB\ZlineBhalf \\[-6.5pt] 
    \ZlineA\ZlineAhalf \\[-6pt] 
    \;\vdots\;\;\vdots\;\,\vdots\;\;\vdots\;\,\vdots\;\,\vdots
  \end{array}
\end{array}.
\nonumber\\[-18pt]
\end{align}
The first two patterns, which are equivalent under the periodic boundary condition in the circular direction,
correspond to the radial stripe pattern of the wild-type zebrafish
in \fgref{fig:Patterns}(a),
while the third one corresponds to the circular stripe pattern in \fgref{fig:Patterns}(b).
This implies that both of the patterns (a) and (b) in \fgref{fig:Patterns}
can be realized as a result of the growth of the retina,
depending on the initial conditions, at $t=0$.
It does not agree with real observations.
In the next section, we will see how it is altered at $t>0$.

\begin{figure}[h]
  \centering
\includegraphics[scale=1]{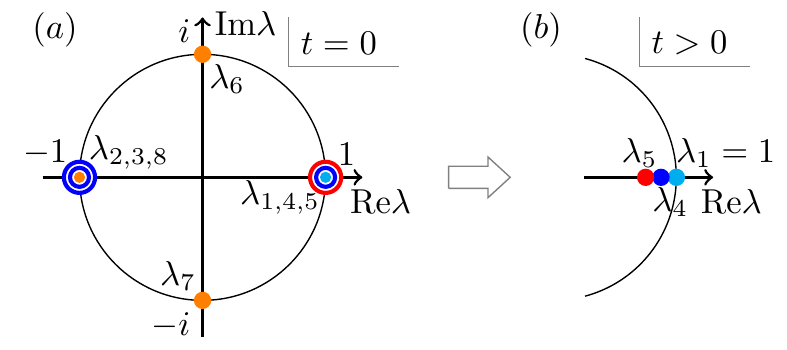}
\vspace{-5pt}
  \caption{
(a): Major (largest 8) eigenvalues $\{\lambda_i^{(0)}\}$ of $\hat{T}_0$ 
and (b): the perturbed eigenvalues $\lambda_{1,4,5}$ for $t>0$.
The cyan and blue points ($\lambda_{1,2,3,4}$) corresponds to the eigenmodes responsible for the radial stripes,
while the red and orange ones ($\lambda_{5,6,7,8}$) are for the circular stripes.
At finite $t$, the cyan $(l=1)$ does not move, 
whereas the others go toward the inner direction.
}
  \label{fig:eigenvalues}
\end{figure}

Corresponding to the multi-stability
due to the reducibility of \eqref{eq:T0},
there arise eight eigenvectors $\{\ket{\lambda_l}\}_{l=1\dots 8}$
with $\abs{\lambda_l}=1$,
and in particular,
there are three with $\lambda_l=1$.
They are manifestly shown in 
\eqref{eq:uv} and \eqref{eq:T=0eigenmodes}
in \supsecref{sup:perturbation}.
The eight eigenvalues are displayed on the complex plane,
in \fgref{fig:eigenvalues}(a).

%---------------------------------
%---------------------------------

%-------------------------------------------------------------
\section{Dynamical pattern selection (DPS) at $t \neq 0 $}
\label{sec:t>0}
%-------------------------------------------------------------

Let us now study DPS at a finite 
effective temperature 
$t>0$.
In this case, the degeneracy of the eigenvalues of $\hat{T}$
is broken by infinitesimal fluctuations.
Apart from $\lambda_1=1$ which is unchanged under the fluctuations,
the other seven major eigenvalues move toward the inside of the unit circle
 $\abs{\lambda_{l=2,\cdots,8}}<1$.
(This is sketched in \fgref{fig:eigenvalues}.
Detailed analyses of  $\{\lambda_l\}$ and $\{\ket{\lambda_l}\}$ at $t>0$
are given in \supsecref{sup:perturbation}.)

To study what happens in this case in a manifest way,
we perform
a stochastic simulation with the transition matrix \eqref{eq:Pxy}.
At each $n$-th column ($n\ge 1$), we start with a random configuration
and sweep through all the sites inside the column 
by $N_\mathrm{sw}$ times using the Metropolis update with $\cE_{ij}$. 
Then we
fix the configuration and go to the $(n+1)$-th column.
This procedure gives a single pattern (a trajectory) 
for each trial starting from an initial configuration at $n=0$.
By repeating this procedure either by changing or keeping the initial condition,
trajectories with the total number $N_\mathrm{traj.}$ are generated.
One may consider each trajectory as a pattern realized in individual zebrafish.
Mean properties of the patterns can be obtained by averaging over the trajectories.
 
%-------------  Fig.3 and Fig.4  ----------------------------- 
\begin{figure}
  \centering
\vspace{-8pt}
\begin{align*}
\begin{array}{l}
(a) \\[-12pt]
\begin{array}{ll}
\hspace{0.012\textwidth}
&\qquad\xrightarrow{\text{\quad Growth\quad}} \\
&\includegraphics[height=0.442\textwidth,angle=90]{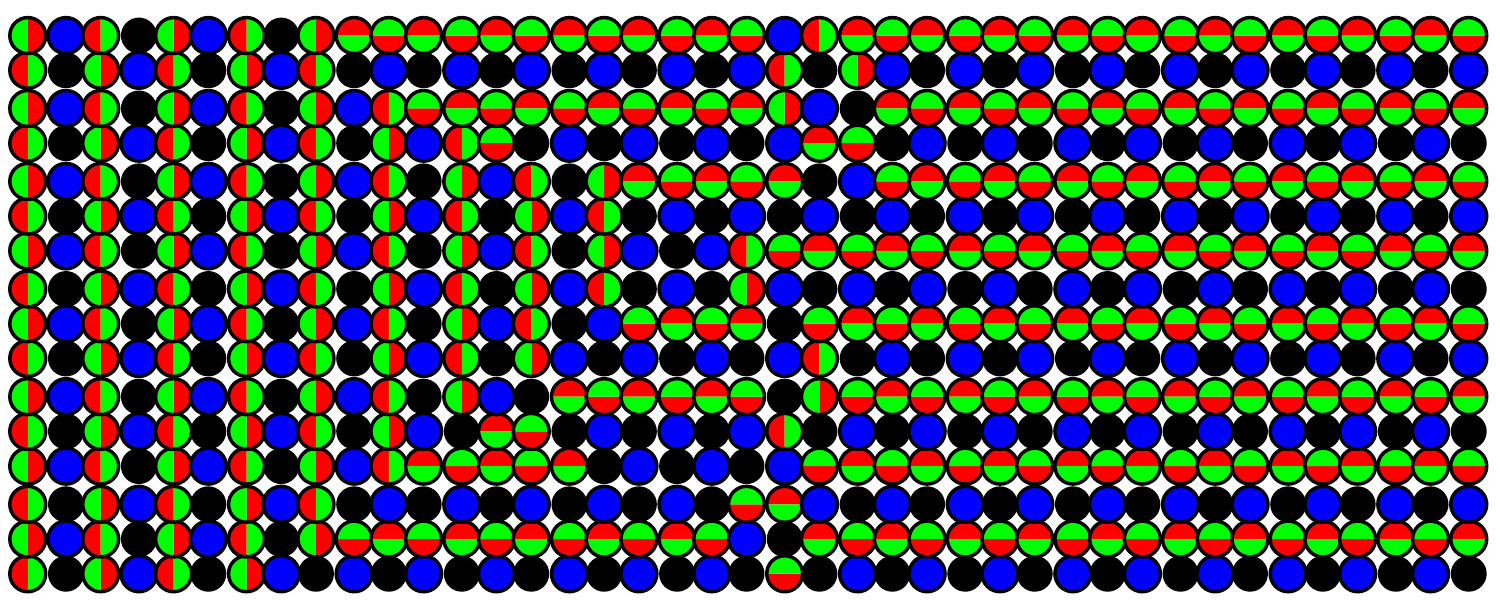} \\
&\includegraphics[height=0.442\textwidth,angle=90]{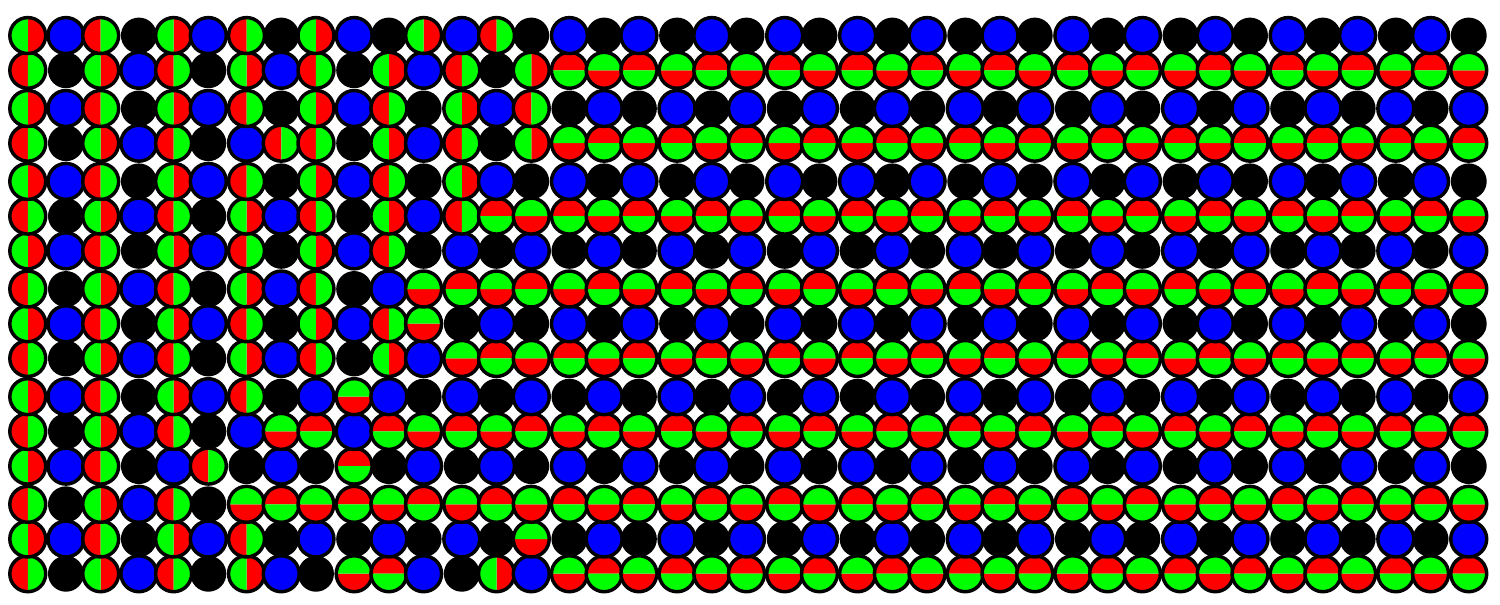}
\end{array}
\\
(b) \\
\includegraphics[width=0.475\textwidth]{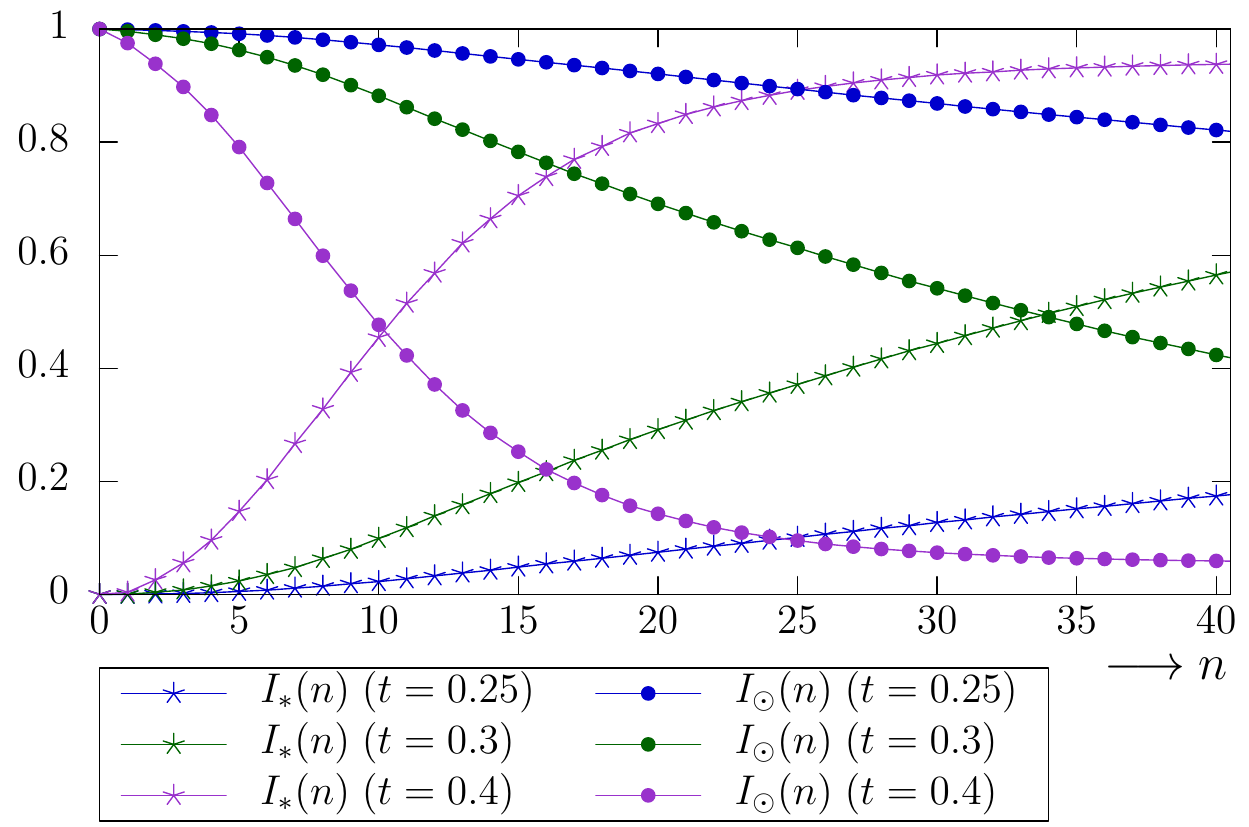}
\end{array}
\end{align*}
\vspace{-20pt}
  \caption{
(a):
Two cases sampled from 2048 trajectories with $F$-type
initial configuration for $w=16$ and $t=0.4$.
(b):
Similarity measures $I_\rad(n)$
(star, at t = 0.25, 0.3, 0.4 from bottom to top curve)
and $I_\cir(n)$
(circle, at t = 0.25, 0.3, 0.4 from top to bottom curve),
averaged over $2048\times 4$ trajectories under
 $(E, F, G, H)$ initial configurations with equal weight.
}
  \label{fig:SimulationFromRotated}
\end{figure}
%-----------------------------------------------------------------

To check whether the circular stripe pattern is indeed unstable,
let us consider the initial configurations of circular-type ($E, F, G, H$).
Shown in \fgref{fig:SimulationFromRotated}(a) are two examples
sampled from $2048$ trajectories with the $F$-type initial configuration at $n=0$
with $w=16$ (about 10\% of the retina circumference) and $N_\mathrm{sw}=2^{17-18}$
(large enough for thermalization).
From the simulations carried out for $t = 0.2 - 0.5$, examples at $t=0.4$ are displayed in the figure:
Although the circular stripe pattern is as well a stationary pattern
as the radial one at $t=0$,
it decays into radial stripes after certain steps
at $t>0$.
Also, a close look at the figure shows that
the transition from the circular stripes to the radial stripes is initiated by the creation and propagation
of small local defects.

\begin{figure}
  \centering
\vspace{-8pt}
\begin{align*}
\begin{array}{l}
(a) \\[-12pt]
\begin{array}{ll}
\hspace{0.012\textwidth}
&\qquad\xrightarrow{\text{\quad Growth\quad}} \\
&\includegraphics[height=0.442\textwidth,angle=90]{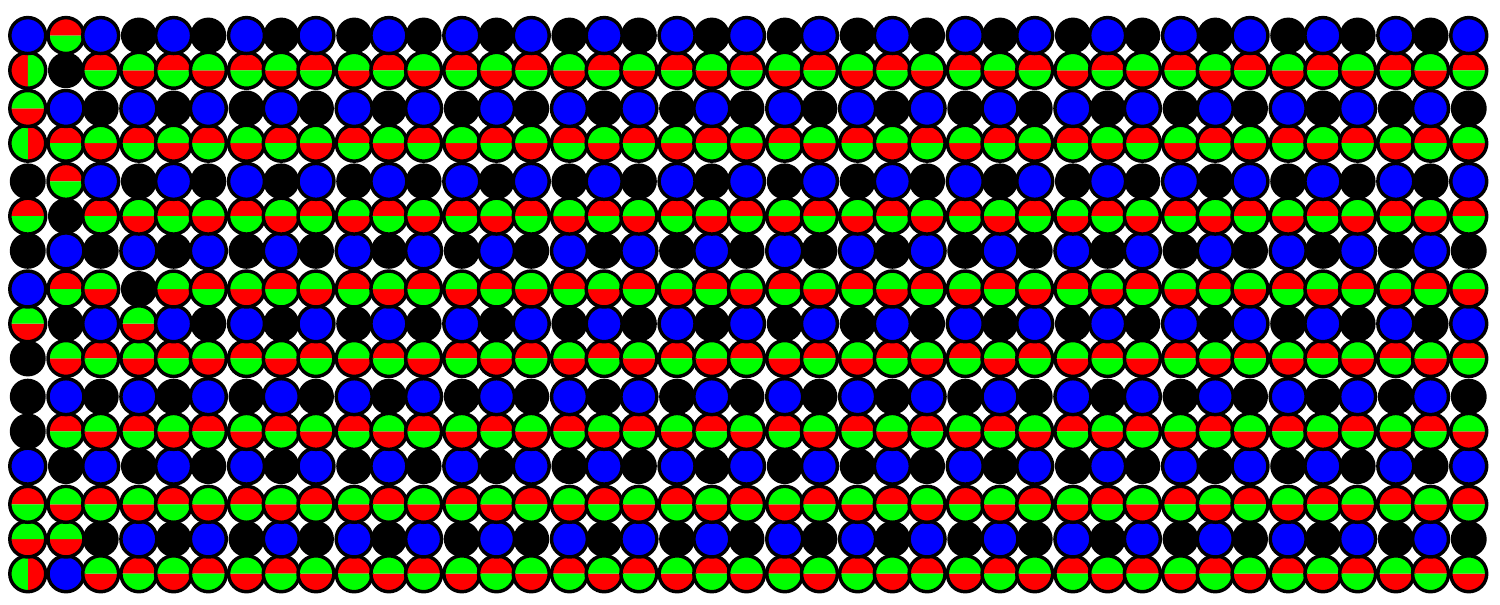} \\
&\includegraphics[height=0.442\textwidth,angle=90]{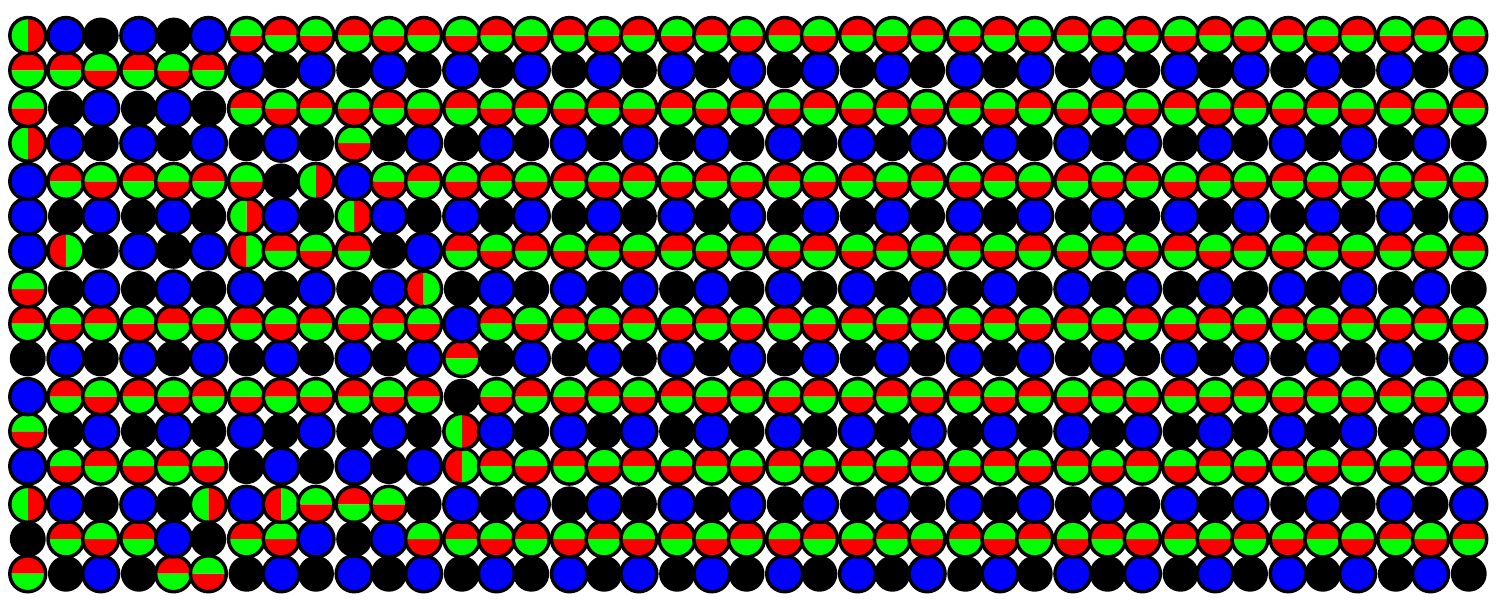}
\end{array}
\\
(b)\\
\includegraphics[width=0.475\textwidth]{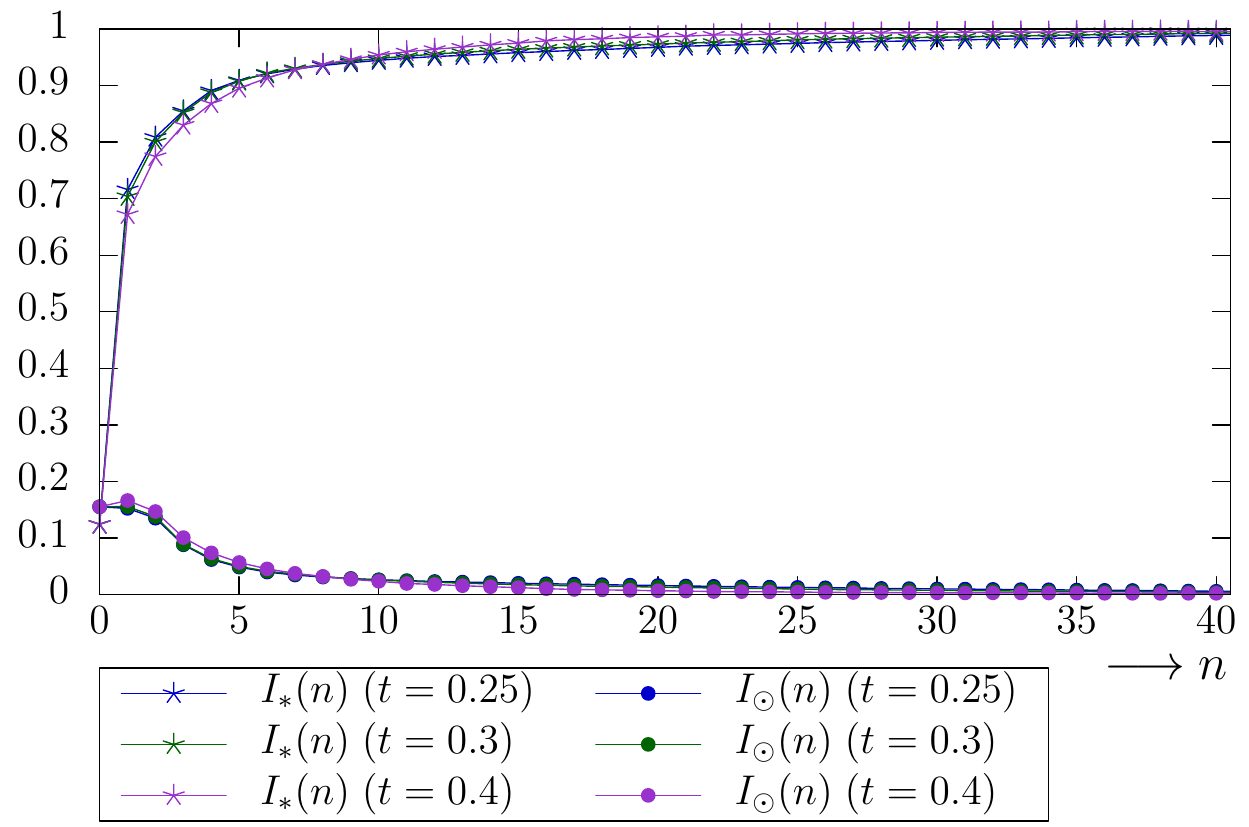}
\end{array}
\end{align*}
\vspace{-20pt}
  \caption{
(a):
Two cases sampled from $4096$ trajectories with random initial
configuration for $w=16$ and $t=0.4$.
(b):
$I_{\rad}(n)$
(star, at t = 0.25, 0.3, 0.4
from top to bottom curve in $n\le 5$
and 
from bottom to top in $n\ge 9$
)
and $I_\cir(n)$
(circle, at t = 0.25, 0.3, 0.4
from top to bottom curve in $n\le 7$
and
from bottom to top in $n\ge 9$),
averaged over $4096$ trajectories.
}
  \label{fig:SimulationFromRandom} 
\end{figure}

In \fgref{fig:SimulationFromRandom},
we show the results of simulations with $w=16$ 
starting from random initial configurations.  This is more realistic and
 is expected to be realized in the wild-type 
zebrafish. Dynamical pattern selection toward the radial stripe pattern
 occurs in qualitatively the  same way as the case of circular-type initial 
 configurations in \fgref{fig:SimulationFromRotated}: 
Qualitatively, $I_\rad(n)$ ($I_\cir(n)$) in the present case approaches to $1$ ($0$) more quickly 
with weaker $t$-dependence.

In order to quantify similarity of a configuration at given $n$ with the radial and circular stripes,
we introduce  a {\it similarity measure}.
First, we note that the radial stripe pattern at given $n$ consists of the basic units
$\{(\uSC\hgrDC)^\top, (\hgrDC\bSC)^\top, (\bSC\hrgDC)^\top, (\hrgDC\uSC)^\top\}$,
while the circular stripe pattern consists of
$\{(\vrgDC\vgrDC)^\top, (\vgrDC\vrgDC)^\top, (\bSC\uSC)^\top, (\uSC\bSC)^\top\}$.
Then, we count how many basic units are contained in a column at given $n$.
Dividing the resultant number by $w$ and averaging it over trajectories,
we obtain the similarity measures $I_\rad(n)$ and $I_\cir(n)$.

Plotted in \fgref{fig:SimulationFromRotated}(b) are the similarity measures 
averaged over
$(E,F,G,H)$-type initial configurations,
as a function of $n$ for $t=0.25, 0.3$ and $0.4$.
One finds that $I_\rad(n)$ ($I_\cir(n)$) approaches to $1$ ($0$) as $n$
increases, which indicates that the radial stripe pattern is the true asymptotic configuration
corresponding to $\ket{P^{(\infty)}}$.
The approach to the radial stripe pattern
becomes slower as $t$ decreases,
as expected by the formation probability of small defects.
The qualitative above features hold also in the realistic situation  with random 
initial configurations, although the approach to the radial stripe pattern is
 achieved faster with less $t$-dependence
(\fgref{fig:SimulationFromRandom}(b)).

All of the results above display fairly good qualitative agreements
with real observations at retina,
where the radial stripe pattern is gradually formed along the growth
and the circular pattern never appears.

Let us finally discuss the essential mechanism of DPS in terms of
the eigenvalues and eigenvectors of $\hat{T}$.
The similarity measure with $R={\rad},{\cir}$ can be written as
\begin{align}
\label{eq:In_expand}
  I_R(n) = \sum_{j=1}^N I_R^{(j)} P_j^{(n)}
  = \sum_{l=1}^N c_l I^{(\lambda_l)}_R \lambda_l^n,
\end{align}
where $I_R^{(j)}$ is the similarity measure of a state $\ket{j}$ to $R$,
and $I_R^{(\lambda_l)} \equiv \sum_j I_R^{(j)} \braket{j}{\lambda_l}$.
For small $t$, $I_{\rad}^{(\lambda_1)}$, $I_{\rad}^{(\lambda_5)}$ and $I_{\cir}^{(\lambda_5)}$ are close to $1$,
while others are close to $0$
(see \supsecref{sup:perturbation}).
This shows that the evolution of the similarity measure along with the growth 
is directly governed by the eigenvectors and eigenvalues of the transition matrix.
For large $n$, we have
$I_{\rad}(n) \rightarrow  I_{\rad}^{(\lambda_1)}  + c_5 I_{\rad}^{(\lambda_5)}  \lambda_5^n \simeq 1$
and $I_{\cir}(n) \rightarrow  c_5 I_{\cir}^{(\lambda_5)} \lambda_5^n \simeq 0$.
Here $\lambda_5=1-\oo(g^2)$ with $g\equiv e^{-1/t}$ corresponds to
the eigenvector $\ket{\lambda_5}\propto (-1,-1,-1,-1,1,1,1,1,\vec{0})^\top+\oo(g^2)$
which swaps the radial and circular stripe patterns.
These are consistent with what are shown in
\fgref{fig:SimulationFromRotated}.
From the figure, one may also extract the ``decay constant'' of the 
circular patterns $I_{\cir} (n) \sim \lambda_5^n = \exp(-n/\bar{n})$ as
$\bar{n}^{-1}= -  \ln \lambda_5  \simeq  (11.6 \pm 1.4) g^2$.
Thus the mean number of steps for the circular stripes to decay is found to be $\bar{n} =13.0\pm 1.6$ for $t=0.4$
(see \supsecref{sup:numerical}).

For DPS to take place, 
the escaping rates from the radial and circular patterns to other patterns in each step play a crucial role.
They are governed by 
the smallest excitation energy to disturb the present pattern in the next column,
which we call
the \textit{minimum deficit energies} for the patterns.
They are computed as
\begin{align}
\label{eq:DeltaE}
  \Delta\cE_*
&= \min_{X\ne B} \{\cE_{XA}\}-\cE_{BA},
\\
  \Delta\cE_\cir
&= \min\l\{\min_{X\ne F} \{\cE_{XE}\}-\cE_{FE}, \min_{X\ne G}\{\cE_{XF}\}-\cE_{GF}\r\},
\nonumber
\end{align}
where $X$ runs over the $(6^w-1)$ configurations
in the next column.
Since the escaping rates behaves as $\sim \exp(-\Delta\cE/t)$,
the configuration having larger $\Delta\cE$ survives at large $n$
(see \supsecref{sup:reduced}).

When $w$ is large and $\sigma/\sigma'$ is not close to $2$,
only {\it local} defects
contribute to $X$ in \eqref{eq:DeltaE}.  Then, 
under the parameter choice in \eqref{eq:paramrange}
and the periodic boundary condition,
we have
\begin{align}
\label{eq:RelationOfDeltaEs}
\Delta\cE_\rad = \sigma > \sigma' = \Delta\cE_\cir.
\end{align}
This implies the radial stripe survives against the
circular stripe.
For $\sigma/\sigma'$ being close to $2$,
{\it global} defects may also contribute.
It is an interesting problem  to be studied in the future.

%------------------------------
\section{Summary and discussion}
\label{sec:discussion}
%------------------------------
In this paper, we have introduced a Markovian lattice model to describe the dynamics
of cone-cell mosaic patterns in growing retina.
With numerical simulations in the $6^w$-dimensional state-space ($w=16$)
together with theoretical analyses of eigenvalues and eigenvectors of the transition matrix $\hat{T}$,
we found that the growing process naturally selects the radial stripe pattern
observed in the wild-type zebrafish against the circular stripe pattern.

A characteristic feature of our model is that
the dynamical pattern selection (DPS) takes place spontaneously
by the successive appearance of small defects which 
 transfer arbitrary patterns to the radial stripes during the growth.
 We note that some additional mechanisms for directional cues,
such as 
concentration gradients of morphogens and cell polarities \cite{GS2012},
may also provide  the origin of the directionality.
Clarifying the relation and the difference between such a mechanism
and DPS would be  an interesting future problem.

In the present study, the cell-cell bindings $\sigma_{pq}$ and the effective temperature $t$
are unknown free  parameters.
To make firm connections of $\sigma_{pq}$ and $t$ to microscopic cell dynamics,
it would be necessary to make quantitative comparisons between our model predictions
and  in vivo  data.
For the theoretical side, realistic numerical simulations with round shaped retina
needs to be carried out.
For the experimental side, more data are needed on the cells in accretion region and
on the defect frequencies with wild-type and mutant embryos.
Direct measurements of the cell-cell bindings using the atomic force microscope \cite{AFM}
would also be useful.

\begin{acknowledgments}
The authors thank
Sachihiro C. Suzuki, Ichiro Masai, Yuko Nishiwaki and Bernold Fiedler
for valuable discussions and comments.
This work was supported by RIKEN iTHES Project
and iTHEMS Program.
The work of NO was supported by RIKEN Special Postdoctoral Researcher
(SPDR) Program.
\end{acknowledgments}

%%%%%%%%%%%%%%%%%%%%%%%%%%%%%%%%%%%%%%%%%%%%%%%%%%%%%%%%%%%%%%
%%%%%%%%%%%%%%%%%%%%%%%%%%%%%%%%%%%%%%%%%%%%%%%%%%%%%%%%%%%%%%
%%%%                    APPENDICES                        %%%%
%%%%%%%%%%%%%%%%%%%%%%%%%%%%%%%%%%%%%%%%%%%%%%%%%%%%%%%%%%%%%%
%%%%%%%%%%%%%%%%%%%%%%%%%%%%%%%%%%%%%%%%%%%%%%%%%%%%%%%%%%%%%%
\appendix
\section{Perturbation theory for the transition matrix}
\label{sup:perturbation}
Here we show a perturbation analysis of the transition matrix
at low temperature $t$ in terms of the expansion parameter,
\begin{align}
  g \equiv e^{-1/t}.
\end{align}
We consider the $6^w\times 6^w$ transition matrix where $w$ is a multiples of 4, i.e., 
$w=4 w_0$ with  $w_0 = 2, 3, 4, \cdots$.  The numerical results
shown in the text correspond to $w=4w_0=16$.
The transition matrix at low $t$ can be expanded by $g$ around 
$\hat{T}_0$ in Eq.\eqref{eq:T0}.  For Eq.\eqref{eq:paramrange} with $\sigma/\sigma'=3/2$, we have
\begin{widetext}
\begin{subequations}
\label{eq:perturbation}
\begin{align}
\hat{T} &= 
\hat{T}_0 + 
\sum_{\alpha=\frac12,1,\frac32,\frac52} g^{\alpha}\hat{T}_{\alpha}+
g^2 \hat{T}_2 + g^3\hat{T}_3 + \oo{(g^{7/2})},
\end{align}
with
\begin{align}
\hat{T}_0&=
\text{\footnotesize\mbox{$
    \left(
    \begin{array}{c|l}
    \begin{array}{cccc|cccc}
    0 & 1 & 0  & 0  &   &   &   &   \\
    1 & 0 & 0  & 0  &   &   &   &   \\
    0 & 0 & 0  & 1  &   & \multicolumn{2}{c}{\hsymb{0}}  &  \\
    0 & 0 & 1  & 0  &   &   &   &   \\
\hline
      &   &   &   & 0 & 0 & 0 & 1   \\
      &   &   &   & 1 & 0 & 0 & 0   \\
      & \multicolumn{2}{c}{\hsymb{0}}  &   & 0 & 1 & 0 & 0 \\
      &   &   &   & 0 & 0 & 1 & 0 \\
    \end{array}
      & \;\hsymb{P_0}\\
      \hline
      & \hspace{30pt}\\
\\
     \hsymb{0} 
      & \;\hsymb{Q_0}
\\
\\
    \end{array}
    \right)
$}},
\qquad
\hat{T}_{\alpha}=
\text{\footnotesize\mbox{$
    \left(
    \begin{array}{c|l}
    \begin{array}{cccc|cccc}
     \;\; & \;\; & \;\;  & \;\; & \;\; & \;\; & \;\; & \;\;\\
     &  &   &   &   &   &   &   \\
      & \multicolumn{2}{c}{\hsymb{0}}  & &   & \multicolumn{2}{c}{\hsymb{0}}  &  \\
     &  &   &   &   &   &   &   \\
\hline
      &   &   &   &  &  &  &    \\
      &   &   &   &  &  &  &    \\
      & \multicolumn{2}{c}{\hsymb{0}}  & &   & \multicolumn{2}{c}{\hsymb{0}} &  \\
      &   &   &   &  &  &  &  
    \end{array}
      & \;\hsymb{P_\alpha}\\
      \hline
      & \hspace{30pt}\\
\\
     \hsymb{0} 
      & \;\hsymb{Q_\alpha}
\\
\\
    \end{array}
    \right)
$}}
\qquad\Big(\alpha=\f{1}{2},1,\f{3}{2},\f{5}{2}\Big)
,
\nonumber\\
\hat{T}_2
&=
\text{\footnotesize\mbox{$
    \left(
    \begin{array}{c|c|l}
& & \\ 
\qquad \hsymb{0}\qquad  & \quad\hsymb{0}\quad & \\
& & \\
\cline{1-2}
\hsymb{0} &
    \begin{array}{cccc}
      0 & 0 & 0 & -\mu_2w_0   \\
      -\mu_1w_0 & 0 & 0 & 0   \\
      0 & -\mu_2w_0 & 0 & 0 \\
      0 & 0 & -\mu_1w_0 & 0
    \end{array}
    & \hsymb{\cdots}
\\
\hline
& & 
\hspace{25pt}
\\
 & & \\
\hsymb{0} & \hsymb{J}& \hsymb{\ddots}\\
 & & \\
    \end{array}
    \right)
$}},
\quad
\nonumber\\
\hat{T}_3
&=
\text{\footnotesize\mbox{$
    \left(
    \begin{array}{c|c|c|l}
\multicolumn{2}{c|}{
    \begin{array}{cccc}
      0 & -\nu w_0 & 0 & 0   \\
      -\nu w_0 & 0 & 0 & 0   \\
      0 & 0 & 0 & -\nu w_0 \\
      0 & 0 & -\nu w_0 & 0
    \end{array}
}
& \quad\hsymb{0}\quad & \\
\cline{1-3}
\multicolumn{2}{c|}{
\hsymb{0}
}
&
    \begin{array}{cccc}
      0 & 0 & 0 & -\xi_2w_0   \\
      -\xi_1w_0 & 0 & 0 & 0   \\
      0 & -\xi_2w_0 & 0 & 0 \\
      0 & 0 & -\xi_1w_0 & 0
    \end{array}
    & \hsymb{\cdots}
\\
\hline
\hspace{40pt} & & & 
\hspace{25pt}
\\
& & \\
\quad  \hsymb{K} \qquad\qquad
&
\hsymb{L} 
& \hsymb{M} & \hsymb{\ddots}\\
& & \\
    \end{array}
    \right)
$}},
\end{align}
\end{subequations}
\end{widetext}
where
$P_0$, $P_\alpha$, $Q_0$, $Q_\alpha$, $J$, $K$, $L$ and $M$ 
are the matrices with the sizes,  
$8 \times (6^w-8)$,  $8 \times (6^w-8)$, 
$(6^w-8) \times (6^w-8)$, $(6^w-8) \times (6^w-8)$,
 $(6^w-8) \times 4$, $(6^w-8)\times 2$, 
 $(6^w-8) \times 2$ and $(6^w-8)\times 4 $, respectively.
With the probability discussed in the main text,
$
\bSC:\uSC:\hgrDC:\vgrDC:\hrgDC:\vrgDC 
= \frac{1}{4} : \frac{1}{4} : \frac{1}{8} : \frac{1}{8} : \frac{1}{8} : \frac{1}{8}
$,
the numerical factors in the above matrices are computed explicitly as
\begin{align}
  \mu_1=8, \quad \mu_2=3,
\quad
  \nu=5,
\quad
  \xi_1=32, \quad \xi_2=6.
\end{align}

For later convenience, we define
\begin{subequations}
\label{eq:PQ}
\begin{align}
\hat{S} &\equiv \hat{T}_0 + \sum_{\alpha=\frac12,1,\frac32,\frac52} g^{\alpha}\hat{T}_{\alpha},
%\nonumber
\\%\quad
{P} &\equiv P_0 + \sum_{\alpha=\frac12,1,\frac32,\frac52} g^{\alpha} P_{\alpha},
%\nonumber
\\
%\quad
{Q} &\equiv Q_0 + \sum_{\alpha=\frac12,1,\frac32,\frac52} g^{\alpha} Q_{\alpha},
\end{align}
\end{subequations}

Note that both $\hat{T}_0$ and $\hat{S}$ are reducible and non-symmetric matrices.
Also they have the same $8 \times 8$ matrix in the upper left corner.
Because  of these properties, they share 8 eigenvalues with $|{\lambda}^{(0)}_{l=1,\cdots,8}|=1$.
We label these eigenvalues and the corresponding left (right) eigenvectors of $\hat{S}$ as
\begin{align}
\label{eq:uv}
&{\lambda}^{(0)}_{l}, 
\quad   \bra{\lambda_l^{(0)}} =\Big(\vec{x}_l,\;\vec{r}_l\Big),
\quad  \ket{\lambda_l^{(0)}} =\Big(\vec{y}_l,\;\vec{0}\Big)^\top
\nonumber\\
&(l=1, \cdots, 8),
\end{align}
with
\begin{widetext}
\begin{subequations}
\label{eq:T=0eigenmodes}
\begin{align}
\lambda^{(0)}_1 &=1, 
& \vec{x}_1 &= (1,1,1,1,1,1,1,1),
& \vec{y}_1 &= \f{1}{4}(1,1,1,1,0,0,0,0),
\\
\lambda^{(0)}_2 &=-1, 
& \vec{x}_2 &= (1,-1,0,0,0,0,0,0),
& \vec{y}_2 &= \f{1}{2}(1,-1,0,0,0,0,0,0),
 \\
\lambda^{(0)}_3 &=-1, 
& \vec{x}_3 &= (0,0,1,-1,0,0,0,0),
& \vec{y}_3 &= \f{1}{2}(0,0,1,-1,0,0,0,0),
\\
\lambda^{(0)}_4 &=1, 
& \vec{x}_4 &= (1,1,-1,-1,0,0,0,0),
& \vec{y}_4 &= \f{1}{4}(1,1,-1,-1,0,0,0,0),
\\
\label{eq:lxy5}
\lambda^{(0)}_5 &=1, 
& \vec{x}_5 &= (0,0,0,0,1,1,1,1),
& \vec{y}_5 &= \f{1}{4}(-1,-1,-1,-1,1,1,1,1),
\\
\lambda^{(0)}_6 &=i, 
& \vec{x}_6 &= (0,0,0,0,1,i,-1,-i),
& \vec{y}_6 &= \f{1}{4}(0,0,0,0,1,-i,-1,i),
\\
\lambda^{(0)}_7 &=-i, 
& \vec{x}_7 &= (0,0,0,0,1,-i,-1,i),
& \vec{y}_7 &= \f{1}{4}(0,0,0,0,1,i,-1,-i),
\\
\lambda^{(0)}_8 &=-1, 
& \vec{x}_8 &= (0,0,0,0,1,-1,1,-1),
& \vec{y}_8 &= \f{1}{4}(0,0,0,0,1,-1,1,-1),
\end{align}
\end{subequations}
\end{widetext}
and the residual vectors,
\begin{align}
\label{eq:ri}
  \vec{r}_l = \vec{x}_l P\big(\lambda^{(0)}_l  -Q\big)^{-1}.
\end{align}
Here  the left and right eigenvectors satisfy
$\braket{\lambda_l^{(0)}}{\lambda_{l'}^{(0)}} = \delta_{l l'}$.
We have chosen particular bases
within the same eigenvalue as shown above for later convenience.
Since the sum of each columns of $Q$ is equal to or smaller than $1$,
the absolute magnitude of each eigenvalue of $Q$ is smaller than $1$, so that
$\big(\lambda^{(0)}_l -Q\big)^{-1}$  in \eqref{eq:ri} always exists.
Also, we have $\vec{r}_1=(1,1,1,\dots)\equiv \vec{1}$.

Now, we consider the effects of the first-order perturbation given by $\hat{T}_{2,3}$ in \eqref{eq:perturbation}.
Since the eigenvalues $+1$ and $-1$ have triple degeneracy, respectively,
we need to diagonalize $\hat{T}_{2,3}$ in each degenerate subspace according to
the degenerate perturbation theory.  
Then we obtain the general form of the eigenvalues,
\begin{align}
 \lambda= \lambda^{(0)} + a g^2 + b  g^3 + \oo(g^4).
\end{align}
Here $a$ and $b$ are written in terms of the matrix elements of $P, Q, J, K, L, M$.
Note that the coefficients $a$ and $b$ depend on $g$, since $P$ and $Q$ are the function of $g$.
In case that we need to obtain a strict power series expansion of $\lambda$ in terms of $g$,
further expansions of $a(g)$ and $b(g)$ should be made.

\subsection[Perturbation in $\lambda_{1,4,5}^{(0)}=1$ subspace]{Perturbation in $\bm{\lambda_{1,4,5}^{(0)}=1}$ subspace}
The eigenspace with $\lambda^{(0)}=1$ in the unperturbed system is
spanned by the $l=1,4,5$ eigenvectors. 
Then,
$\hat{T}_2$ is expressed by a $3\times 3$ matrix as
\begin{align}
\bra{\lambda_l^{(0)}}\hat{T}_2\ket{l'^{(0)}}_{(l,l')\in \{1,4,5\}^2}
&=
  \begin{pmatrix}
    0 & 0 & 0 \\
    0 & 0 & 0 \\
    0 & 0 & \f{1}{4}\vec{r}_5\cdot\vec{j} - \f{1}{2}({\mu_1+\mu_2})w_0
  \end{pmatrix}
%\quad
\nonumber\\
\text{with}
\quad
  \vec{j} &\equiv J\begin{pmatrix}1\\1\\1\\1\end{pmatrix}.
\end{align}
where we used
$\vec{r}_4\cdot\vec{j}=0$ obtained from the symmetry of the 1-cell shift.
(Due to the periodic boundary condition, our model 
is invariant under the 1-cell shift of each column,
e.g.  $(\uSC\hrgDC\vgrDC\bSC\uSC\bSC\hgrDC\vrgDC)^\top \to (\vrgDC\uSC\hrgDC\vgrDC\bSC\uSC\bSC\hgrDC)^\top$.
With this transformation, $\vec{j}$ is trivially invariant,
whereas $\vec{r}_4$ changes its sign.
Therefore, the parity-odd quantity such as $\vec{r}_4\cdot\vec{j}$ vanishes.)

Together with the -perturbation by $\hat{T}_3$,
\begin{align}
  \bra{\lambda_5^{(0)}}\hat{T}_3\ket{\lambda_5^{(0)}}
  &= -\l(\nu+\f{1}{2}(\xi_1+\xi_2)\r)w_0
\nonumber\\
  &\qquad + \f{1}{4}\vec{r}_5\cdot(-\vec{k}+\vec{\ell}+\vec{m}),
\end{align}
with $\vec{k}$, $\vec{\ell}$ and $\vec{m}$ being the sum of the columns in 
$K$, $L$ and $M$ respectively, 
we obtain
\begin{align}
\label{eq:perturb-lambda5}
  \lambda_5 &= 1 + \l(-\f{11}{2}w_0+\f{1}{4}\vec{r}_5\cdot\vec{j}\r)g^2
\nonumber\\
  &\quad+
  \l\{-24w_0 + \f{1}{4}\vec{r}_5\cdot(-\vec{k}-\vec{\ell}+\vec{m})\r\}g^3
  + \oo(g^4).
\end{align}
  
The degeneracy between $\lambda_1$ and $\lambda_4$ even under the influence of $\hat{T}_2$
is removed by $\hat{T}_3$ represented as 
\begin{align}
&\bra{\lambda_l^{(0)}}\hat{T}_3\ket{\lambda_{l'}^{(0)}}_{(l,l')\in\{1,4\}^2}
\nonumber\\
&\quad=
  \begin{pmatrix}
\f{1}{4}\vec{1}\cdot(\vec{k}+\vec{\ell}) - \nu w_0 & 
\f{1}{4}\vec{r}_4\cdot(\vec{k}+\vec{\ell}) \\
\f{1}{4}\vec{1}\cdot(\vec{k}-\vec{\ell}) & 
\f{1}{4}\vec{r}_4\cdot(\vec{k}-\vec{\ell}) - \nu w_0 \\
  \end{pmatrix}
\nonumber\\
&\quad=
  \begin{pmatrix}
0 & 0 \\
0 & 
\f{1}{4}\vec{r}_4\cdot(\vec{k}-\vec{\ell}) - \nu w_0 \\
  \end{pmatrix}.
\end{align}
This leads to 
\begin{align}
\label{eq:perturb-lambda14}
  \lambda_1 = 1, \qquad
  \lambda_4 = 1 - \Big\{5w_0 - \f{1}{4}\vec{r}_4\cdot\big(\vec{k}-\vec{\ell}\big)\Big\}\,g^3 + \oo(g^4). 
\end{align}

The major eight eigenvalues for $t=0$ \eqref{eq:T=0eigenmodes}
and the perturbed $\lambda_{1,4,5}$ for small positive $t$
\eqref{eq:perturb-lambda5}\eqref{eq:perturb-lambda14}
are sketched in \fgref{fig:eigenvalues}.

\subsection[Perturbation in $\lambda_{2,3,8}^{(0)}=-1$ subspace]{Perturbation in $\bm{\lambda_{2,3,8}^{(0)}=-1}$ subspace}
The $\lambda^{(0)}=-1$ eigenspace in the unperturbed system is 
spanned by the $l=2,3,8$ eigenvectors,
\begin{align}
\bra{\lambda_l^{(0)}}\hat{T}_2\ket{\lambda_{l'}^{(0)}}_{(l,l')\in \{2,3,8\}^2}
&=
  \begin{pmatrix}
    0 & 0 & 0 \\
    0 & 0 & 0 \\
    0 & 0 & \f{1}{2}(\mu_1+\mu_2)w_0 + \f{1}{4}\vec{r}_8\cdot\vec{j}_8
  \end{pmatrix}
%\quad
\nonumber\\
&\text{with}
\quad
  \vec{j}_8 \equiv J\begin{pmatrix}1\\-1\\1\\-1\end{pmatrix}.
\end{align}
Together with the perturbation by $\hat{T}_3$,
\begin{align}
 \bra{\lambda_8^{(0)}}\hat{T}_3\ket{\lambda_8^{(0)}}&= \f{\xi_1+\xi_2}{2}w_0 + \f{1}{4}\vec{r}_8\cdot\vec{m}_8, 
%\quad
\nonumber\\
\text{with}
\quad
  \vec{m}_8 &\equiv M\begin{pmatrix}1\\-1\\1\\-1\end{pmatrix}, \\
\bra{l^{(0)}}\hat{T}_3\ket{l'^{(0)}}_{(l,l')\in \{2,3\}^2}
&=
  \begin{pmatrix}
    \nu w_0 + \f{1}{2}\vec{r}_2\cdot\vec{k}_- 
  & \f{1}{2}\vec{r}_2\cdot\vec{\ell}_- \\
    \f{1}{2}\vec{r}_3\cdot\vec{k}_- 
  & \nu w_0 + \f{1}{2}\vec{r}_3\cdot\vec{\ell}_-
  \end{pmatrix},
\end{align}  
we have 
\begin{widetext}
\begin{align}
\label{eq:perturb-lambda8}
  \lambda_8 &= -1 + \Big(\f{11}{2}w_0+\f{1}{4}\vec{r}_8\cdot\vec{j}_8\Big)\,g^2 
%\nonumber\\
%&\qquad
+ \Big(19w_0+\f{1}{4}\vec{r}_8\cdot\vec{m}_8\Big)\,g^3 
+ \oo(g^4), 
\end{align}
and 
\begin{align}
\label{eq:perturb-lambda23}
\lambda_{2,3}= -1 +
\bigg\{
\nu w_0 + \f{1}{4}(\vec{r}_2\cdot\vec{k}_- + \vec{r}_3\cdot\vec{\ell}_-)
%\nonumber\\
\pm \f{1}{4}\s{(\vec{r}_2\cdot\vec{k}_- - \vec{r}_3\cdot\vec{\ell}_-)^2 + 4(\vec{r}_2\cdot\vec{\ell}_-)(\vec{r}_3\cdot\vec{k}_-)}\bigg\} g^3
+ \oo(g^4).
\end{align}
\end{widetext}

\subsection[Perturbation for $\lambda_{6,7}^{(0)}=\pm i$]{Perturbation for $\bm{\lambda_{6,7}^{(0)}=\pm i}$}
The $\oo(g^2)$ and $\oo(g^3)$ perturbations on $\lambda_6$ are obtained from
\begin{subequations}
\begin{align}
\bra{\lambda_6^{(0)}}\hat{T}_2\ket{\lambda_6^{(0)}}
  &= -\f{1}{2}(\mu_1+\mu_2)w_0i + \f{1}{4}\vec{r}_6\cdot\vec{j}_6,
%\qquad
\nonumber\\
\bra{\lambda_6^{(0)}}\hat{T}_3\ket{\lambda_6^{(0)}}
  &= -\f{1}{2}(\xi_1+\xi_2)w_0i + \f{1}{4}\vec{r}_6\cdot\vec{m}_6,
\end{align}
where
\begin{align}
  \vec{m}_6 \equiv M\begin{pmatrix}1\\-i\\-1\\i\end{pmatrix},
\qquad
  \vec{j}_6 \equiv J\begin{pmatrix}1\\-i\\-1\\i\end{pmatrix}.
\end{align}
\end{subequations}
One then  finds
\begin{align}
\label{eq:perturb-lambda67}
  \lambda_6 
=& i 
+ \Big(-\f{11}{2}w_0i + \f{1}{4}\vec{r}_6\cdot\vec{j}_6\Big)g^2
\nonumber\\
&+ \Big(-19w_0i + \f{1}{4}\vec{r}_6\cdot\vec{m}_6\Big)g^3
+ \oo(g^4),
%\qquad
\\
\lambda_7 =& (\lambda_6)^*.
\nonumber
\end{align}

These eigenvalues
\eqref{eq:perturb-lambda5}, \eqref{eq:perturb-lambda14},
\eqref{eq:perturb-lambda8}, \eqref{eq:perturb-lambda23}
and \eqref{eq:perturb-lambda67}
reveal 
the behaviors of this Markovian lattice model at low temperature.
First, the stationary distribution after a long sequence generated by 
Eq.\eqref{eq:Markov} is given by the $\lambda=1$ eigenstate, $\ket{1}$.
At the leading order,
it is solely composed of the radial stripes,
which directly shows the dynamical pattern selection.
The $\oo(g^2)$-perturbation \eqref{eq:perturb-lambda5} implies that
the circular stripes decay to the radial stripes
rather quickly with a lifetime of $\oo(g^{-2})$.
The shifts of the radial stripes
(from ABAB\dots to CDCD\dots or BABA\dots, for example),
corresponding to the decay of the modes of
$\ket{\lambda_2}$, $\ket{\lambda_3}$ or $\ket{\lambda_4}$,
are much slower than that, with a lifetime of $\oo(g^{-3})$.

\subsection{Perturbation for eigenvectors}

As well as the eigenvalues discussed above,
the corresponding eigenvectors also receive perturbation, 
\begin{align}
  \ket{\lambda_l} &= \ket{\lambda_l^{(0)}} + g^2\ket{\lambda_l^{(2)}} + g^3\ket{\lambda_l^{(3)}} + \oo(g^4),
\nonumber\\
  &\ket{\lambda_l^{(2)}} = 
\sum_{\lambda_m^{(0)}\ne\lambda_l^{(0)}}\f{\bra{\lambda_m^{(0)}}\hat{T}_2\ket{\lambda_l^{(0)}}}{\lambda_l^{(0)}-\lambda_m^{(0)}}\;\ket{\lambda_m^{(0)}},
%\qquad
\nonumber\\
  &\ket{\lambda_l^{(3)}} = 
\sum_{\lambda_m^{(0)}\ne\lambda_l^{(0)}}\f{\bra{\lambda_m^{(0)}}\hat{T}_3\ket{\lambda_l^{(0)}}}{\lambda_l^{(0)}-\lambda_m^{(0)}}\;\ket{\lambda_m^{(0)}}.
\end{align}
In particular, for $l=1,2,3,4$,
we have $\hat{T}_2 \ket{\lambda_l^{(0)}} =0$ leading to $\ket{\lambda_l^{(2)}}=0$. 
Because the similarity measures $I_R^{(\lambda_l)}$ depend linearly on the state vectors,
we obtain
\begin{subequations}
\label{eq:PEV}
\begin{align}
  I_{\rad}^{(\lambda_l)}&=
  \begin{cases}
    1 + \oo(g^3) & (l=1) \\
    \oo(g^3) & (l=2,3,4) \\
    -1+\oo(g^2) & (l=5)  \\
    \oo(g^2) & (l=6,7,8)
  \end{cases},
\\
%\qquad
  I_{\cir}^{(\lambda_l)}&=
  \begin{cases}
    \oo(g^3) & (l=1,2,3,4)\\
    1+\oo(g^2) & (l=5) \\
    \oo(g^2) & (l=6,7,8)
  \end{cases}.
\end{align}
\end{subequations}

\section{Numerical values of perturbative coefficients}
\label{sup:numerical}
\subsection{Analytic formulas for $w=8$}
For $w=4w_0=8$, we can explicitly compute the matrices
$P$, $Q$, $J$, $K$, $L$, $M$ and the resulting perturbative coefficients.
The formula \eqref{eq:ri} 
for the ``unperturbed'' eigenvectors (including the effects up to $\hat{T}_{3/2}$)
is expanded as
\begin{align}
  \vec{r}_l = \vec{x}_l\f{{P}}{\lambda^{(0)}_l} \sum_{n=0}^{\infty}  \l(\f{{Q}}{\lambda^{(0)}_l}\r)^n,
\end{align}
from which we can calculate
the coefficients of $\vec{r}_l$ expanded in terms of $g$
with a pretty good convergence by taking the sum up to $n=128$.
Substituting them into 
\eqref{eq:perturb-lambda5}, \eqref{eq:perturb-lambda14},
\eqref{eq:perturb-lambda8}, \eqref{eq:perturb-lambda23}
and \eqref{eq:perturb-lambda67},
we finally obtain the perturbative coefficients for $\lambda_l$ as
\begin{widetext}
\begin{subequations}
\label{eq:lambdaperturb}
\begin{align}
  \lambda_1&= 1,\\
  \lambda_{2,3}&= -1 + \oo(g^4), \\
  \lambda_4&= 1 - 10 g^3 + \oo(g^4), \\
\label{eq:lambda5perturb}
  \lambda_5&= 1 - 5.8 g^2 - 16.84 g^{5/2}+ 9.068 g^3 -371.5176\dot{7}4\dot{0} g^{7/2}+ \oo(g^4), \\
  \lambda_{6,7}&= \pm \Big(1 - 5.8 g^2 - 16.84 g^{5/2} + 9.068 g^3 
+ 324.1043\dot{4}0\dot{7} g^{7/2}\Big)i
- 47.41\dot{3} g^{7/2} + \oo(g^4), \\
  \lambda_8&= -1 + 5.8 g^2 + 16.84 g^{5/2} -9.068 g^3 + 276.6910\dot{0}7\dot{4} g^{7/2} + \oo(g^4),
\end{align}
\end{subequations}
\end{widetext}
where the dots represent recurring decimals.
The true coefficients are known to be rational numbers, 
so that the above coefficients must be exact although  obtained numerically.
We find that $|\lambda_{5,6,7,8}|$ are degenerated up to $\oo(g^3)$,
and the $\oo(g^{7/2})$ contributions start to break the degeneracy.
Also, $\lambda_{6,7}$ acquire nonzero real parts from $\oo(g^{7/2})$.
We note that $\lambda_{2,3}$ are not perturbed from $-1$ up to $\oo(g^{7/2})$:
Their absolute values should eventually become smaller than 1 in higher orders
due to the Perron-Frobenius theorem.

\subsection{Numerical results for $w=8$ }
%---  Figs ------------
\begin{figure}
  \centering
  \includegraphics[width=0.48\textwidth]{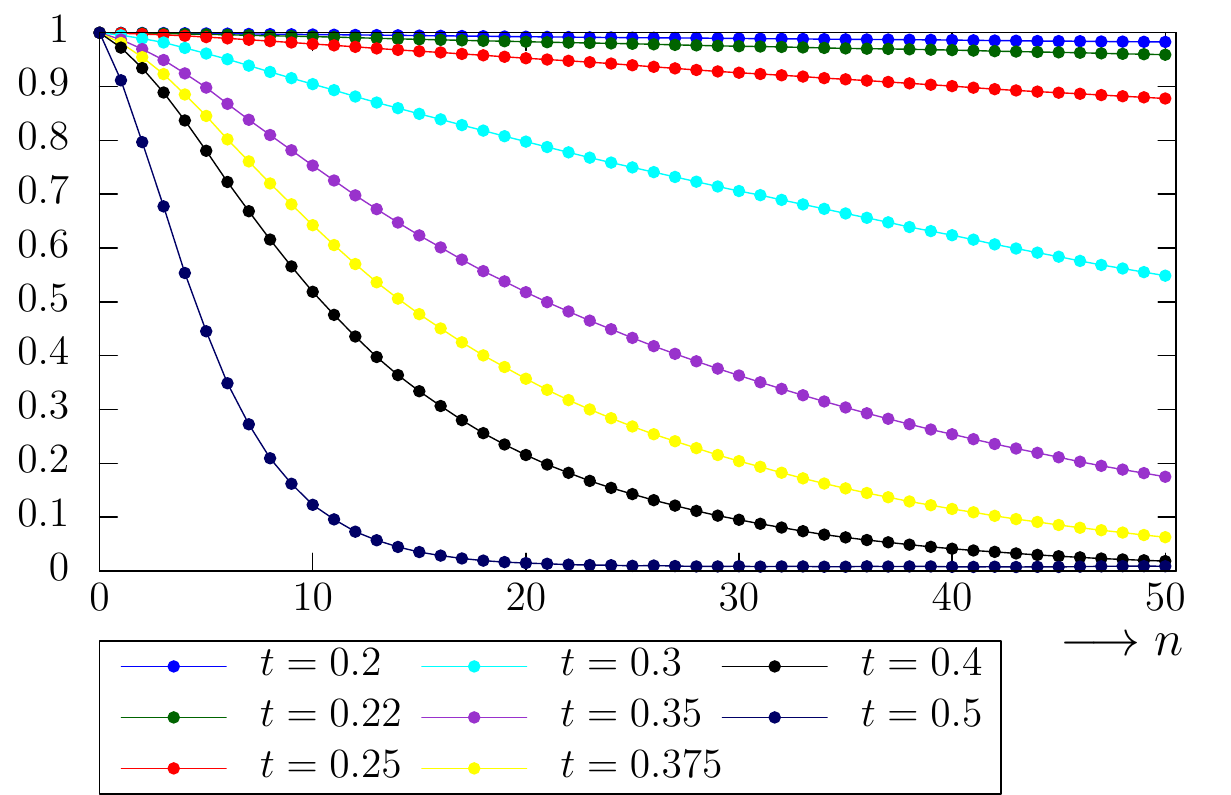}
  \caption{Plot of $I_\cir(n)$ 
(at t = 0.2, 0.22, 0.25, 0.3, 0.35, 0.375, 0.4, 0.5
from top to bottom curve)
by simulations for $w=8$,
averaged over $4096\times 4$ trajectories 
under
 $(E, F, G, H)$ initial configurations with equal weight
for each temperature.
}
  \label{fig:w=8InPlot}
\end{figure}
%--------------------------

The analytic result \eqref{eq:lambda5perturb}
can be checked from the numerical result of $I_{\cir}(n)$ as a function of $n$. 
Shown in \fgref{fig:w=8InPlot} are the results of simulations for $4096\times 4$ trajectories
with equal-weight $(E, F, G, H)$ initial configurations.
Corresponding initial state at $n=0$ is written as
\begin{align}
  \sum_l c_l\ket{\lambda_l} = \ket{\lambda_1^{(0)}} + \ket{\lambda_5^{(0)}},
\end{align}
which leads to the coefficient $\{c_l\}$ as
\begin{align}
\label{eq:cl}
  c_1&=1,
\quad
  c_4=\oo(g^4),
\quad
  c_5=1+\oo(g^4),
%\quad
\nonumber\\
  c_l &= -\f{\bra{\lambda_l^{(0)}}\hat{T}_2\ket{\lambda_5^{(0)}}}{1-\lambda_l^{(0)}}g^2 + \oo(g^3)
\quad (l\ne 1,4,5).
\end{align}

From \eqref{eq:In_expand} together with \eqref{eq:PEV} and \eqref{eq:cl},
we have
\begin{align}
\label{eq:lambda5dominates}
  I_\cir(n) \simeq \lambda_5^n I_\cir^{(\lambda_5)},
\end{align}
in a range of $n$ satisfying 
$  1 \ll n \ll -g^{-2}\log{g}$.
Therefore, we obtain
\begin{align}
 \lambda_5^{\rm eff} \simeq  \f{I_{\cir}(n+1)}{I_{\cir}(n)}. 
 \label{eq:ana-num}
\end{align}
Plotted in \fgref{fig:fitting}(a) is a comparison of the 
analytic result of $\lambda_5$ in \eqref{eq:lambda5perturb}
and the numerical result of  $\lambda_5^{\rm eff}$ in \eqref{eq:ana-num}.
They show good agreement at low temperature ($t\le 0.25$), while the the discrepancy appears
for $t \ge 0.3$ indicating the breakdown of the perturbative estimate at high temperature.
By fitting the numerical data at $t=0.2, 0.22, 0.25$ with an ansatz,
$\lambda_5 = 1 - \gamma g^2 - \delta g^{5/2}$,
we obtain
\begin{align}
  \gamma \simeq 5.78 \pm 1.82,
\qquad
  \delta \simeq 20.2 \pm 10.5,
\end{align}
which are quite consistent with analytic results in \eqref{eq:lambda5perturb};
$(\gamma, \delta)=(5.8,16.84)$.

\subsection{Numerical results for $w=16$}
\begin{figure}
  \centering
\begin{tabular}{l}
\quad (a)
\\[-5pt]
  \includegraphics[height=0.45\textwidth]{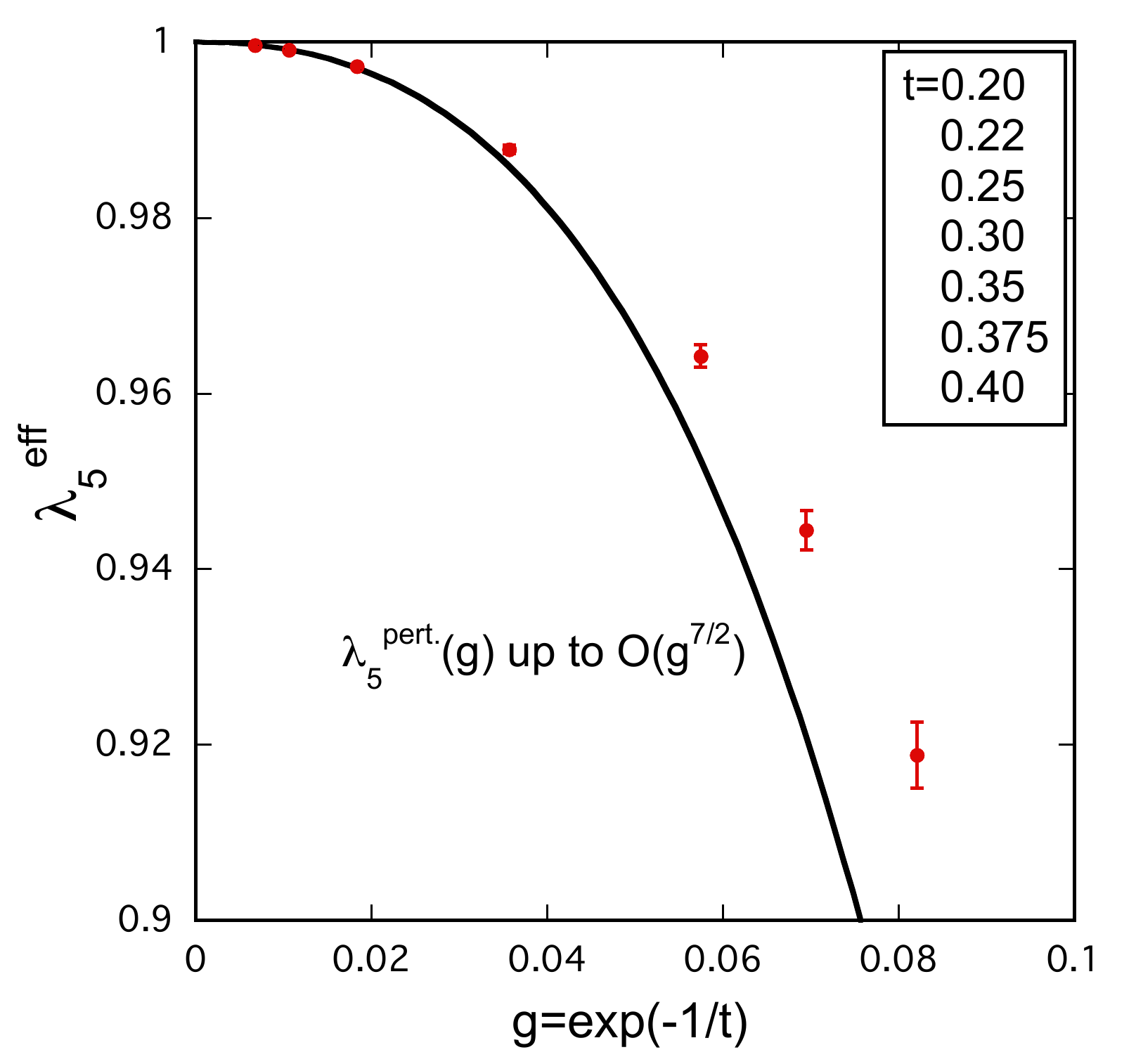}
\\
\quad (b)
\\[-5pt]
  \includegraphics[height=0.45\textwidth]{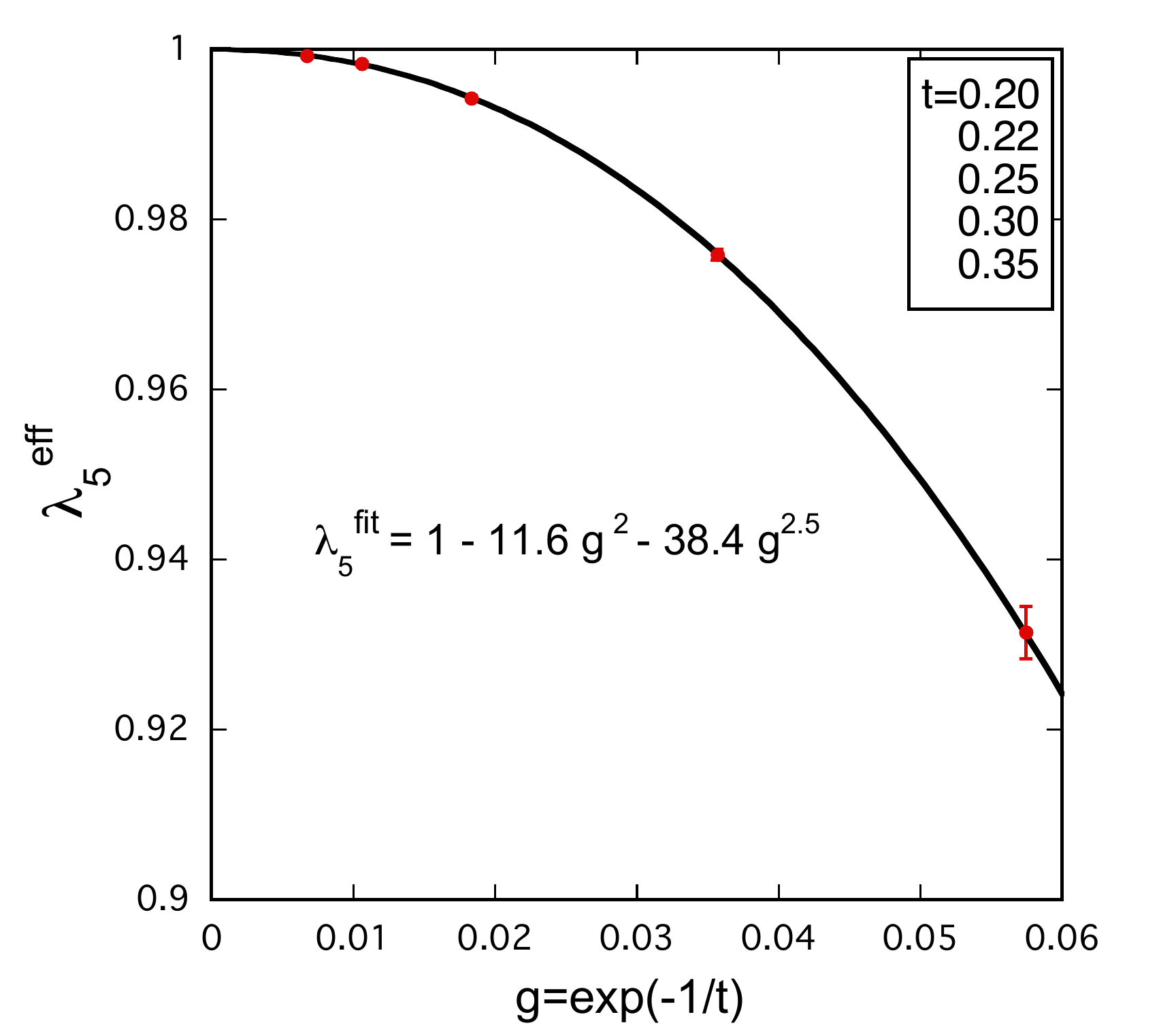}
\end{tabular}
  \caption{
Ranges of $I_{\cir}(n+1)/I_{\cir}(n)$ for $10\le n\le 30$ (red marks)
and 
the perturbative $\lambda_5$ up to $g^{7/2}$-order
\eqref{eq:lambda5perturb} (black curve), 
(a) for $w=8$ from the data of \fgref{fig:w=8InPlot} 
and
(b) for $w=16$ from that of \fgref{fig:SimulationFromRotated}.
}
  \label{fig:fitting}
\end{figure}

For $w=4w_0=16$, it is numerically demanding to obtain the analytic formula similar
to \eqref{eq:lambdaperturb}.
Nevertheless, we can extract the coefficients $\gamma$ and $\delta$
by plotting  $\lambda_5^{\rm eff}$ in \eqref{eq:ana-num} as shown in \fgref{fig:fitting}(b). 
By fitting the data  at $t=0.2, 0.22, 0.25, 0.30, 0.35$, 
we find
\begin{align}
\label{eq:w16-gd}
  \gamma\simeq 11.6 \pm 1.4,
\qquad
  \delta\simeq 38.4 \pm 8.0.
\end{align}
These results together with \eqref{eq:lambda5dominates} lead to 
the temperature dependence of the decay rate of the
circular stripe pattern as given in the main text.

%------------------------------
\section{Mechanism of DPS in a Minimal Model}
\label{sup:reduced}
%------------------------------

To identify the essential origin of the dynamical pattern selection (DPS),
let us consider a minimal toy-model with only 3 states (rather than $6^w$ states discussed in the main text).
They represent two stable patterns (which we call $X$ and $Y$)
and a transition state ($Z$).
They have intra energies $B^{\rm (intra)}_{X,Y,Z}$ and and the inter
binding energies $B^{\rm (inter)}_{YX,ZX,ZY}$ and obey the Markov chain with  Eq.\eqref{eq:Pxy}.
The transition matrix at zero temperature ($t=0$) is given as
\begin{align}
&\quad\, {\scriptstyle X\;\; Y\;\; Z}
\nonumber\\
  \hat{T}_0
=&
  \begin{pmatrix}
    1 & 0 & p_1\\
    0 & 1 & p_2\\
    0 & 0 & q
  \end{pmatrix},
\end{align}
where $p_{1,2}$ and $q$ are positive with the constraint $p_1+p_2+q=1$.
The eigenvalues of this matrix is $\lambda^{(0)}=1, q$,
and we define the corresponding basis vectors as
\begin{align}
  \bra{\lambda_1} &= \l(1,0,\f{p_1}{p_1 + p_2}\r),
&
  \ket{\lambda_1} &=
  \begin{pmatrix}
    1 \\ 0\\ 0
  \end{pmatrix};
\nonumber\\
%\quad
   \bra{\lambda_2} &= \l(0,1,\f{p_2}{p_1 + p_2}\r),
&
   \ket{\lambda_2} &=
  \begin{pmatrix}
    0 \\ 1\\ 0
  \end{pmatrix};
%\quad
\nonumber\\
   \bra{\lambda_3} &= \l(0,0,1\r),
&
   \ket{\lambda_3} &=
  \begin{pmatrix}
    -\f{p_1}{1-q} \\ -\f{p_2}{1-q} \\ 1
  \end{pmatrix}.  
\end{align}

For small nonzero temperature, $0 < t \ll 1$,
we perturb the transition matrix  as
$\hat{T} = \hat{T}_0 + \delta \hat{T}$
by
\begin{align}
\label{eq:toy-T}
 \delta  \hat{T}=
  \begin{pmatrix}
    -a-a' & b'    & c \\
    a'    & -b-b' & d \\
    a & b & e
  \end{pmatrix},
\end{align}
where
$a, b, a', b' > 0$. 
In terms of the binding energies,

\begin{align}
\label{eq:ab}
  a 
&= \f{p_{_X} e^{-\beta\cE_{ZX}}}{p_{_X} e^{-\beta\cE_{XX}} + p_{_Y} e^{-\beta\cE_{YX}} + p_{_Z} e^{-\beta\cE_{ZX}}}
\sim e^{-\beta(\cE_{ZX}-\cE_{XX})},
\nonumber\\
  b 
&= \f{p_{_X} e^{-\beta\cE_{ZY}}}{p_{_X} e^{-\beta\cE_{XY}} + p_{_Y} e^{-\beta\cE_{YY}} + p_{_Z} e^{-\beta\cE_{ZY}}}
\sim e^{-\beta(\cE_{ZY}-\cE_{YY})},
\nonumber\\
  a' 
&= \f{p_{_X} e^{-\beta\cE_{YX}}}{p_{_X} e^{-\beta\cE_{XX}} + p_{_Y} e^{-\beta\cE_{YX}} + p_{_Z} e^{-\beta\cE_{ZX}}}
\sim e^{-\beta(\cE_{YX}-\cE_{XX})},
\nonumber\\
  b' 
&= \f{p_{_X} e^{-\beta\cE_{ZY}}}{p_{_X} e^{-\beta\cE_{XY}} + p_{_Y} e^{-\beta\cE_{YY}} + p_{_Z} e^{-\beta\cE_{ZY}}}
\sim e^{-\beta(\cE_{ZY}-\cE_{YY})}.
\end{align}
where $\cE_{ij}$'s are defined in the same  way as Eq.\eqref{eq:Pxy}.

Projected on the $\lambda^{(0)}=1$ unperturbed eigenspace, 
$\delta \hat{T}$ becomes
\begin{align}
\bra{\lambda_l^{(0)}} \delta\hat{T}\ket{\lambda_{l'}^{(0)}}_{(l,l')\in\{1,2\}^2}
=
  \begin{pmatrix}
    -\f{p_2}{p_1+p_2}a - a' & \f{p_1}{p_1+p_2}b + b' \\
    \f{p_2}{p_1+p_2}a + a'  & -\f{p_1}{p_1+p_2}b - b'
  \end{pmatrix},
\end{align}
whose eigenvalues read
\begin{align}
\label{eq:3states-deltalambda}
  \delta \lambda_1 = 0, \quad 
  \delta \lambda_2 =  -(a+b+a'+b').
\end{align}  

Corresponding eigenvectors in the original 3-dimensional space are
\begin{align}
\label{eq:toy-sp2}
  \ket{\lambda_1}
= 
\begin{pmatrix}
    p_1b + (p_1+p_2)b'\\
    p_2a + (p_1+p_2)a'\\
    0
\end{pmatrix},
\qquad
   \ket{\lambda_2}
   = 
\begin{pmatrix}
    -1\\1\\0
\end{pmatrix}.
\end{align}
Then in the first-order perturbation theory,
$\ket{\lambda_1}$ 
is the eigenvector of $\hat{T}$ for the eigenvalue $\lambda=1$,
i.e.,
it represents the stationary distribution of the system.

When $p_1$ and $p_2$ are of the same order,
which one of $X$ and $Y$ dominates the steady distribution is determined by
which of $\max\{a,a'\}$ and $\max\{b,b'\}$
is larger than the other.
From \eqref{eq:ab}, 
the magnitude of $\max\{a,a'\}$ ($\max\{b,b'\}$)
corresponds to the minimum excitation energy from $X$ ($Y$).
If  we define
\begin{align}
  \Delta\cE_X &= \min\{\cE_{ZX},\cE_{YX}\} - \cE_{XX},
%\quad
\nonumber\\
  \Delta\cE_Y &= \min\{\cE_{ZY},\cE_{XY}\} - \cE_{YY},
\end{align}
then $\Delta\cE_X > \Delta\cE_Y$ ($\Delta\cE_X < \Delta\cE_Y$) indicates that
$\ket{\lambda_1} \simeq (1,0,0)^\top$ [$\ket{\lambda_1} \simeq (0,1,0)^\top$], 
i.e. the pattern $X$ ($Y$) is dynamically selected.
Also, the mean-number of steps $\bar{n}$ for the unstable pattern to decay   
is derived from \eqref{eq:3states-deltalambda} as
 \begin{align}
 \bar{n}^{-1} = -\ln\lambda_2 \sim  e^{-\beta \min\{\Delta\cE_X, \Delta\cE_Y\}}.
\end{align}

Although we analyzed the minimal 3-state model here,
it is straightforward to generalize it to
a full-scale model with a block structure,
\begin{align}
  \hat{T}_0
&=
  \begin{pmatrix}
    S_1 & 0  & P_1\\
    0  & S_2 & P_2\\
    0  &  0  & Q
  \end{pmatrix},
\end{align}
which includes the case of the $6^w$-state model discussed in the main text and in
\eqref{eq:perturbation}.
In this context,
$X$ ($Y$) in the minimal model corresponds to the radial stripes $\{A, B, C, D \}$
(circular stripes $\{E, F, G, H \}$).
Also $\ket{\lambda_1} $ ($\ket{\lambda_2}$) in the minimal model corresponds to 
$\ket{\lambda_1} $ in Eq.\eqref{eq:lxy5} ($\ket{\lambda_5}$  in Eq.\eqref{eq:lxy5}).
Then, the stability of the radial stripes is a result of
\begin{align}
\Delta\cE_\rad=\sigma > \sigma'= \Delta\cE_\cir.
\end{align}

%--------- REFERENCES ------------

\end{document}